\documentclass[aps,prd,amsmath,amssymb]{revtex4-1}

\usepackage{graphicx}
\usepackage{natbib}
\usepackage{subcaption}

\begin{document}

\title{Excited Boson Stars}

\author{Lucas G. Collodel}
% \email{lucas.gardai.collodel@uni-oldenburg.de}
\author{Burkhard Kleihaus}

 \author{Jutta Kunz}

\affiliation{
 Institut fur Physik, Universitat Oldenburg, Postfach 2503 D-26111 Oldenburg, Germany}

\begin{abstract}
Axisymmetric rotating radially excited boson stars are analyzed. 
For several fixed parameter sets, the full sets of solutions are obtained. 
In contrast to the nodeless boson stars, the radially excited
sets of solutions do not exhibit a spiraling behavior.
Instead they form a loop, when the boson frequency is varied
from its maximal value given by the boson mass to a minimal value and back.
Thus for all allowed boundary data of the scalar field at the origin,
there are two distinct solutions, except for the endpoints.
While one endpoint corresponds to the trivial solution,
the other one represents the most compact solution.
The energy density and the pressures of the solutions are analyzed.
A decomposition of the scalar field into spherical harmonics is performed,
and stability considerations are presented.
\end{abstract}

\maketitle

\section{Introduction}
First obtained almost half a century ago 
\cite{PhysRev.168.1445,PhysRev.172.1331,PhysRev.187.1767},
boson stars have been studied in many different
theoretical and astrophysical contexts (for reviews see 
e.g.~\cite{LEE1992251,JETZER1992163,doi10.1142/S0218271892000057,Mielke1997re,0264-9381-20-20-201, Liebling2012}).
Boson stars arise, when a complex scalar field is coupled to
gravity, with the simplest boson stars containing only a mass term
but no self-interaction. The inclusion of a repulsive self-interaction
allows for more massive stars \cite{PhysRevLett.57.2485},
while a sextic potential allows for very massive and highly compact
objects, close to the black hole limit \cite{PhysRevD.35.3658}.

The astrophysical applications of boson stars range from 
black hole mimickers 
with respect to different astrophysical scales in terms of the
mass and the accretion rate 
\cite{PhysRevD.80.084023,PhysRevD.62.104012}
all the way to the description of galactic halos and their
lensing properties \cite{PhysRevD.66.023503,doi:10.1111/j.1365-2966.2006.10330.x}.
Also multi-state boson stars have been considered,
which superpose fundamental and excited boson star solutions 
to obtain more realistic rotation curves of spiral galaxies
\cite{PhysRevD.81.044031,PhysRevD.82.123535}.

In boson stars, the complex scalar field
has a harmonic time dependence with frequency $\omega_s$.
Besides the fundamental solutions there are also radially
excited solutions, where the scalar field has any number of nodes.
The properties of spherically symmetric excited boson
stars are very similar to those of the fundmental boson stars.
In particular, they exist between a maximal value of the
boson frequency $\omega_s$, corresponding to the mass of the
boson field, and a minimal value, depending on the 
potential and the coupling to gravity.
Close to the minimal value of the boson frequency $\omega_s$
a spiralling (or oscillating) pattern of the solutions is observed, 
well-known from other compact objects like neutron stars.

When boson stars are set into rotation, regularity imposes
a quantization condition on the scalar field configuration.
Consequently, boson stars can only possess angular momenta $J$
which are integer multiples of their particle number $Q$,
$J=\hbar mQ$ \cite{SCHUNCK1998389}.
The properties of rotating boson stars have been investigated
widely \cite{SCHUNCK1998389,PhysRevD.55.6081,PhysRevD.56.762,
Schunck1999,MIELKE2000185,PhysRevD.72.064002,PhysRevD.77.064025,PhysRevD.85.024045}.
Rotating boson stars follow the pattern of non-rotating boson stars,
when the nodeless solutions are considered, independent of
the quantum number $m$, or the symmetry of the boson field
\cite{PhysRevD.72.064002,PhysRevD.77.064025}. Interestingly,
when rotating sufficiently fast, even ergoregions may arise
for boson stars \cite{PhysRevD.77.124044,PhysRevD.77.064025}.

Motivated by previous studies of multi-state boson stars,
which considered non-rotating boson stars only
\cite{PhysRevD.81.044031,PhysRevD.82.123535},
we have undertaken the study of excited rotating boson stars.
While we have expected to observe the same pattern for
the solutions as seen for the nodeless boson stars,
this expectation has not been born out by our investigations.
Radially excited rotating boson stars do not exhibit a spiralling
or oscillating behavior. Instead for every value of the
frequency in the possible range, there are at least two solutions,
i.e., the sets of rotating excited boson stars evolve from 
the maximal frequency to a minimal frequency and then again all the
way back to the maximal value.

In this paper we construct numerous sets of excited boson stars
and investigate their physical properties. We show that these stars 
must be constituted by bosons in a superposition of modes.
Different mode arrangements allow for equilibrium
configurations and therefore more than one solution is found for a fixed 
value of the field's radial derivative at the origin. 

In Sec.~\ref{s1} we present the model and derive the equations 
we need to solve for. Sec.~\ref{s2}  
contains the description of our numerical methods.
In Sec.~\ref{s3} we present and discuss the solutions.
We conclude in Sec.~\ref{s4}.
The appendices contain the field equations 
and the \emph{vierbein} components used in the diagonalization
of the energy momentum tensor.
The metric signature is taken to be $(-,+,+,+)$, 
Greek indices represent coordinate indices, while Latin indices
are used for the \emph{vierbein} basis. 
We use geometrical units such that $c=\hbar=8\pi G=1$.

\section{Model}
\label{s1}
\subsection{Action, Metric and Ansatz}
\label{s1ss1}
We consider a self-interacting complex scalar field 
that is minimally coupled to gravity,
\begin{equation}
\label{action}
S=\int \left[\frac{R}{2}-\frac{1}{2}g^{\mu\nu}\left(\Phi_{, \mu}\Phi^{*}_{, \nu}+\Phi_{, \nu}\Phi^{*}_{, \mu}\right)-U\left(\lvert\Phi\rvert\right)\right]\sqrt{-g}d^4x, 
\end{equation}
where $R$ is the curvature scalar, $\Phi$ is the complex scalar field, 
$U$ is the self-interaction potential, 
and we employ the compact notation 
$\Phi_{,\mu}\equiv\partial_{\mu}\Phi$. 

To obtain stationary axisymmetric solutions of rotating boson stars, 
we adopt the quasi-isotropic Lewis-Papapetrou metric 
in adapted spherical coordinates $(t,r,\theta,\varphi)$.
The line element then reads
\begin{equation}
\label{metric}
ds^2=-fdt^2+\frac{l}{f}\left[g\left(dr^2+r^2d\theta^2\right)+r^2\sin^2\theta\left(d\varphi-\frac{\omega}{r}dt\right)^2\right].
\end{equation}
This line element contains four scalar functions,
namely $f$, $l$, $g$ and $\omega$. 
%for which we need to solve together with the boson field. 
For axisymmetric solutions these metric coefficients 
must be functions of the radial coordinate $r$ and the polar angle $\theta$, 
only. The spacetime then possesses two Killing vector fields,
\begin{equation}
\label{kvectors}
\xi=\partial_t,\qquad\eta=\partial_\varphi.
\end{equation}

Although the boson field is complex, all of the terms 
in the action are real and functions of $r$ and $\theta$ alone. 
Therefore, any dependence on the coordinates $t$ and $\varphi$ 
is included in a phase factor. 
This results in the ansatz \cite{Schunck1996}
\begin{equation}
\label{ansatz}
\Phi(t,r,\theta,\varphi)=\phi(r,\theta)e^{i\omega_st+im\varphi},
\end{equation}
where $\phi(r,\theta)$ is a real function,
and $\omega_s$ is the frequency of the boson field, 
which is also a real number.
The identification $\Phi(\varphi)=\Phi(\varphi+2\pi)$ requires $m$ 
to be an integer number, which may be considered as the field's 
rotational quantum number. 
%One could just as well take another ansatz by making a parity transformation, 
%$i\rightarrow -i$ on the formula above. This would correspond to an 
%antiboson spinning in the opposite direction. 

The Lagrangian density is invariant under a global $U(1)$ phase transformation 
$\Phi\rightarrow\Phi e^{i\alpha}$, leading to a conserved current
\begin{equation}
\label{current}
j^{\mu}=-i(\Phi^{\ast}\partial^{\mu}\Phi-\Phi\partial^{\mu}\Phi^{\ast}), \qquad \nabla_{\mu}j^{\mu}=0.
\end{equation}

\subsection{Potential}
\label{s1ss2}
In this work, the boson star is a gravitationally bound system 
of massive interacting bosons described by the complex scalar field $\Phi$. 
The sixtic potential
\begin{align}
\label{sixpot}
U\left(\lvert\Phi\rvert\right) &= \lambda\Phi^{\ast}\Phi\left[\left(\Phi^{\ast}\Phi\right)^2-a\Phi^{\ast}\Phi+b\right] \nonumber\\
& = \lambda\phi^2\left(\phi^4-a\phi^2+b\right),
\end{align}
was first proposed as a mechanism for geons to mimic nonlinear spinor fields  
\cite{PhysRevD.20.1303,PhysRevD.24.2111}. 
Friedberg, Lee and Pang then considered 
non-rotating bosonic nontopological solitons
in the presence of gravity, i.e., spherically symmetric boson stars 
\cite{Friedberg:1986tq,LEE1992251}.
A systematic study of nodeless rotating boson stars 
governed by this potential was performed in 
\cite{PhysRevD.72.064002,PhysRevD.77.064025}. 

\begin{figure}[!h]
\centering
\includegraphics[scale=0.7]{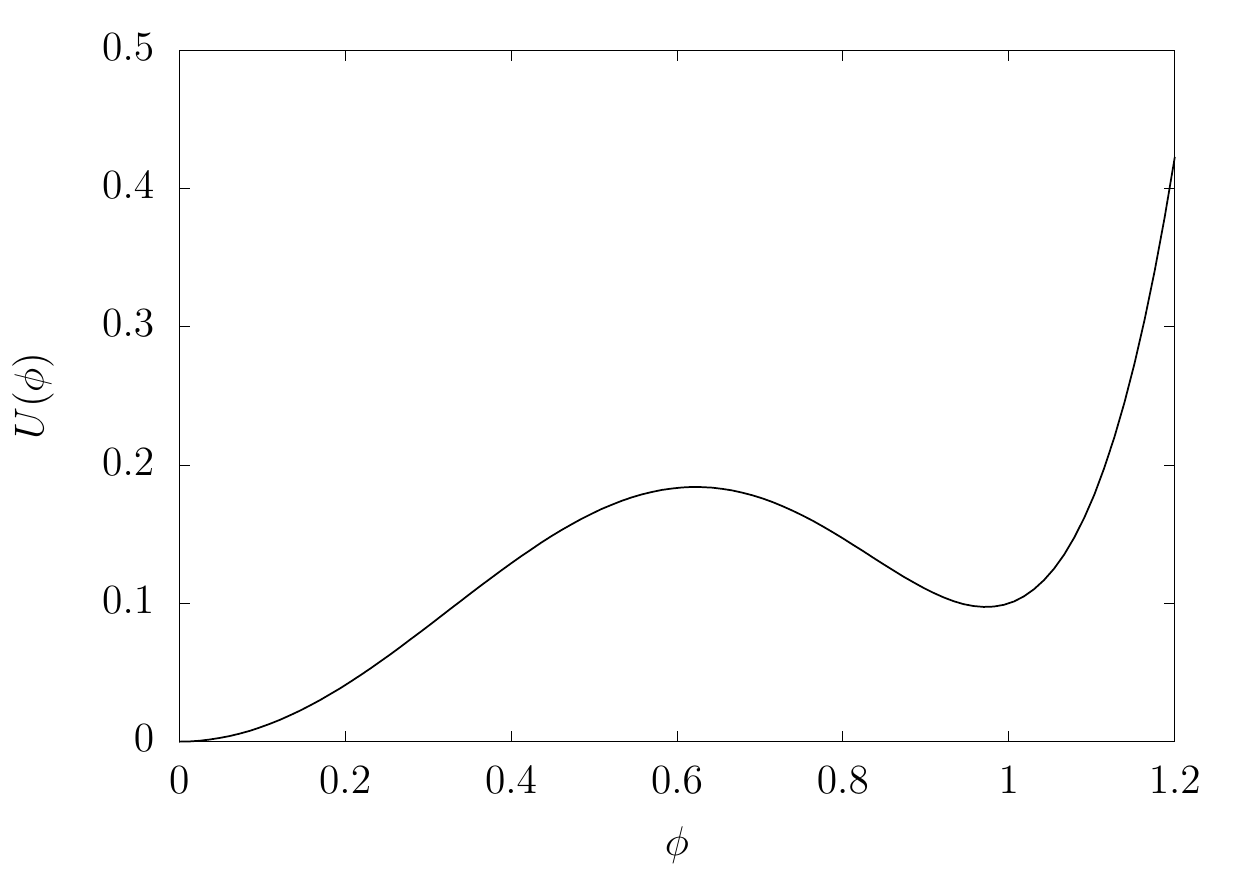}
\caption{The potential $U(\phi)$ for $\lambda=1$, $a=2$, $b=1.1$. 
Note the global minimum at the origin and a local minimum
around $\phi=0.97$.}
\label{potu}
\end{figure}

The stability of the boson stars with a sextic potential was investigated 
via catastrophe theory for spherically symmetric solutions 
\cite{PhysRevD.43.3895} and rotating solutions \cite{PhysRevD.85.024045}. 
The sextic potential is shown in Fig.~\ref{potu}, 
where we see that its parameters can be chosen in such a manner 
that it features a local minimum at a finite
value of $\phi$, besides the global one at $\phi=0$. 
From the quadratic term of the potential we read off
the boson mass, $m_b=\sqrt{\lambda b}$.

Bound solutions exist only for a limited range of the frequency $\omega_s$. 
The lower bound depends on the strength of the gravitational coupling,
on the potential parameters, and on the character of the solution. 
In contrast, the upper bound depends on the boson mass alone
\cite{Friedberg:1986tq,PhysRevD.66.085003, PhysRevD.72.064002}
\begin{equation}
\label{maxomega}
\omega_s^2<\omega_{max}^2\equiv\frac{1}{2}\frac{\partial^2U\left(\phi\right)}{\partial\phi^2}\rvert_{\phi=0}=\lambda b.
\end{equation}
Localized solutions can (in principle) be found for any value of $\omega_s$ 
arbitrarily close to $\omega_{max}$, but at the equality 
the only possible solution is the trivial one. 
This holds both for solitons in Minkowski space
and boson stars in an asymptotically flat spacetime.

\subsection{Equations}
 
Variation of the action (\ref{action}) with respect to the metric yields 
the Einstein field equations
\begin{equation}
\label{efe}
R_{\mu\nu}-\frac{1}{2}Rg_{\mu\nu}=T_{\mu\nu},
\end{equation}
where the energy-momentum tensor $T_{\mu\nu}$ is given by
\begin{align}
T_{\mu\nu} &= g_{\mu\nu}\mathcal{L}_{\Phi}-2\frac{\partial L_{\Phi}}{\partial g^{\mu\nu}} \nonumber \\
           &= -g_{\mu\nu}\left[\frac{1}{2}g^{\rho\tau}\left(\Phi^{\ast}_{, \rho}\Phi_{, \tau}+\Phi^{\ast}_{, \tau}\Phi_{, \rho}\right)+U\left(\lvert\Phi\rvert\right)\right]+\left(\Phi^{\ast}_{, \mu}\Phi_{, \nu}+\Phi^{\ast}_{, \nu}\Phi_{, \mu}\right).
\end{align}

From the above equations it is clear that one can rescale the field $\phi$ 
with a parameter to be varied. For instance, 
\begin{equation}
\kappa T_{\mu\nu}\left(\hat{\phi}\right)=T_{\mu\nu}\left(\phi\right), \qquad \hat{\phi}\equiv\frac{\phi}{\sqrt{\kappa}},
\end{equation}
and $\kappa$ is now a free parameter that takes different values 
to designate either a different gravitational coupling strength 
or different parameters of the potential $U$.
The field equations (\ref{efe}) then take the more general form 
\begin{equation}
\label{efe2a}
R_{\mu\nu}-\frac{1}{2}Rg_{\mu\nu}=\kappa T_{\mu\nu}.
\end{equation}

Variation with respect to the field gives rise to the Einstein-Klein-Gordon equation
\begin{equation}
\label{ekge}
\left(\Box-\frac{\partial U}{\partial\lvert\Phi\rvert^2}\right)\Phi=0,
\end{equation}
where $\Box$ is the covariant D'Alembert differential operator.

\subsection{Boundary Conditions}

The Einstein field equations (\ref{efe2a}) give four independent coupled 
second order elliptic partial differential equations (PDEs) for the metric functions,
which can be put in a diagonal form, such that they assume the canonical form 
of elliptic PDEs with no terms with $\partial^2_{r\theta}$ occurring  (see Appendix \ref{fea}). 
These equations together with eq.~(\ref{ekge}) for the boson field
must all be solved simultaneously. 

In order to do so we are required to establish boundary conditions for all five functions,
that guarantee globally regular bounded solutions. 
Because of the ansatz (\ref{ansatz}) adopted for the scalar field 
and the quadratic dependence of its Lagrangian density $\mathcal{L}_{\Phi}=\mathcal{L}_{\Phi}\left(\Phi^{\ast}\Phi\right)$,
the scalar field can have either even or odd parity 
with respect to reflections on the equatorial plane. 
Due to this symmetry, the angular domain is reduced to $\theta\in\left[0,\pi/2\right]$. 
In this work we will analyze only even parity solutions. 

For regularity at the origin we demand the conditions
\begin{equation}
\partial_rf\rvert_{r=0}=0,\qquad \partial_rl\rvert_{r=0}=0,\qquad g\rvert_{r=0}=1,\qquad \omega\rvert_{r=0}=0, \qquad \phi\rvert_{r=0}=0.
\end{equation}
In contrast to the non-rotating case, where the scalar field takes a constant value at the origin, 
rotation induces a singular point at the origin which is regular only if the scalar field vanishes
at this point, as seen in eq.~(\ref{eomphi}). 
This creates a profound change for the scalar field, which assumes a torus-like shape,
analogous to the hydrogen wave function, 
when the magnetic quantum number is different from zero. 
Near the origin, the scalar field for the rotating boson star has the following expansion
\begin{equation}
\label{phiorigin}
\phi(r,\theta)=\phi_0r^m\sin^m\theta,
\ \ \ \ 
\left.\frac{\partial^m \phi}{\partial r^m}\right|_{\pi/2} = m! \phi_0 .
\end{equation}
The solutions of boson stars will then be parameterized by the value $\phi_0$
for rotating and non-rotating boson stars alike.

The gravitationally bound solutions are required to approach asymptotically 
the Minkowski spacetime. The boundary conditions at infinity read 
\begin{equation}
f\rvert_{r\rightarrow\infty}=1,\qquad l\rvert_{r\rightarrow\infty}=1,\qquad g\rvert_{r\rightarrow\infty}=1,\qquad \omega\rvert_{r\rightarrow\infty}=0, \qquad \phi\rvert_{r\rightarrow\infty}=0.
\end{equation}

Along the symmetry axis the boundary conditions are given by
\begin{equation}
\partial_{\theta}f\rvert_{\theta=0}=0,\qquad \partial_{\theta}l\rvert_{\theta=0}=0,\qquad g\rvert_{\theta=0}=1,\qquad \partial_{\theta}\omega\rvert_{\theta=0}=0, \qquad \phi\rvert_{\theta=0}=0,
\end{equation}
and in the equatorial plane by
\begin{equation}
\partial_{\theta}f\rvert_{\theta=\pi/2}=0,\qquad \partial_{\theta}l\rvert_{\theta=\pi/2}=0,\qquad \partial_{\theta}g\rvert_{\theta=\pi/2}=0,\qquad \partial_{\theta}\omega\rvert_{\theta=\pi/2}=0, \qquad \partial_{\theta}\phi\rvert_{\theta=\pi/2}=0.
\end{equation}
The elementary flatness condition
\begin{equation}
\frac{X_{, \mu}X^{, \mu}}{4X}=1, \qquad X=\eta^{\mu}\eta_{\mu},
\end{equation}
requires
\begin{equation}
g\rvert_{\theta=0}=g\rvert_{\theta=\pi}=1.
\end{equation}

\subsection{Conserved Quantities}

In a stationary asymptotically flat spacetime, the Komar integrals provide a way 
to calculate the global charges associated with the Killing vectors. The mass
\begin{equation}
M=2\int_\Sigma R_{\mu\nu}n^\mu\xi^\nu dV,
\end{equation} 
and the angular momentum
\begin{equation}
J=-\int_\Sigma R_{\mu\nu}n^\mu\eta^\nu dV,
\end{equation} 
are then obtained from an integral bounded at spatial infinity. 
Here, $R_{\mu\nu}$ is the Ricci tensor, $\Sigma$ is a spacelike asymptotically flat hypersurface,
$n^\mu$ is the a vector normal to $\Sigma$ with $n^\mu n_\mu=-1$. 
The metric (\ref{metric}) implies that 
$n^{\mu}=(\xi^\mu+ \eta^\mu \omega/r)/\sqrt{f}$.
Substituting the Ricci tensor by the energy-momentum using eq.~(\ref{efe2a}), 
and noticing that the volume element is $dV=\sqrt{-g/f}drd\theta d\varphi$ one arrives at
\begin{equation}
\label{intm}
M=\int(2T^t_t-T)\sqrt{-g}drd\theta d\varphi, 
\end{equation}
and
\begin{equation}
\label{intj}
J=-\int T^t_\varphi\sqrt{-g}drd\theta d\varphi.
\end{equation}

The mass and angular momentum, being global observables measured at infinity, can also be 
obtained directly from the asymptotic expansion of the solutions \cite{PhysRevLett.86.3704},
\begin{equation}
\label{asympmj}
M=\frac{1}{2}\lim_{r\rightarrow\infty}r^2\partial_rf, \qquad J=\frac{1}{2}\lim_{r\rightarrow\infty}r^2\omega.
\end{equation}

The Noether charge, associated with the particle number $Q$ of the solutions
can be calculated by integrating the projection of the conserved current eq.~(\ref{current}) 
onto the future directed hypersurface normal $n^\mu$ over the whole space,
\begin{align}
\label{intq}
Q &= \int_\Sigma  j_\mu n^\mu dV \nonumber \\
&=\int\frac{2\phi^2}{f}\left(\omega_s+\frac{m\omega}{r}\right)\sqrt{-g}drd\theta d\varphi.
\end{align}
The integrand of the above equation is simply $-j^t\sqrt{-g}$. 
As noted by Schunck and Mielke \cite{Schunck1996}, $T^t_\varphi=mj^t$, 
then comparison of eq.~(\ref{intq}) with eq.~(\ref{intj}) leads to
\begin{equation}
J=mQ,
\end{equation}
and the angular momentum of boson stars is thus quantized.

It is convenient to define a normalization factor for each solution by
\begin{equation}
\label{normalfactor}
N^2=\int_\Sigma\Phi^{\ast}\Phi dV.
\end{equation}

\section{Numerical Approach} \label{s2}

The set of five coupled nonlinear elliptic PDEs (\ref{efe1}-\ref{eomphi}) 
is solved simultaneously, subject to the boundary conditions given above. 
In order to find the solutions satisfying
the appropriate asymptotic behavior, 
we define a compactified radial coordinate
\begin{equation}
\bar{r}=\frac{r}{1+r},
\end{equation}
which goes to one as $r$ goes to infinity. 

The equations are solved using the program 
package FIDISOL \cite{Schoenauer:1989}.
They are discretized on a nonuniform rectangular grid 
in $\bar{r}$ and $\theta$, $[0,1]\times[0,\pi/2]$. 
Then the damped Newton scheme is applied 
until the desired accuracy is obtained. 

Whereas often a known analytic solution of a closely related problem 
can be successfully used as an initial guess, this is, however, 
not possible for these boson stars. 
We therefore proceed by constructing Q-ball solutions 
for a boson frequency $\omega_s$ near $\omega_{\max}$,
and use these as initial guesses for the boson stars, 
since at these values of $\omega_s$ we are approaching 
the trivial solution, where Q-balls and boson stars tend to be very similar.

\begin{figure}[h!]
\begin{subfigure}{0.4\textwidth}
\includegraphics[scale=0.58]{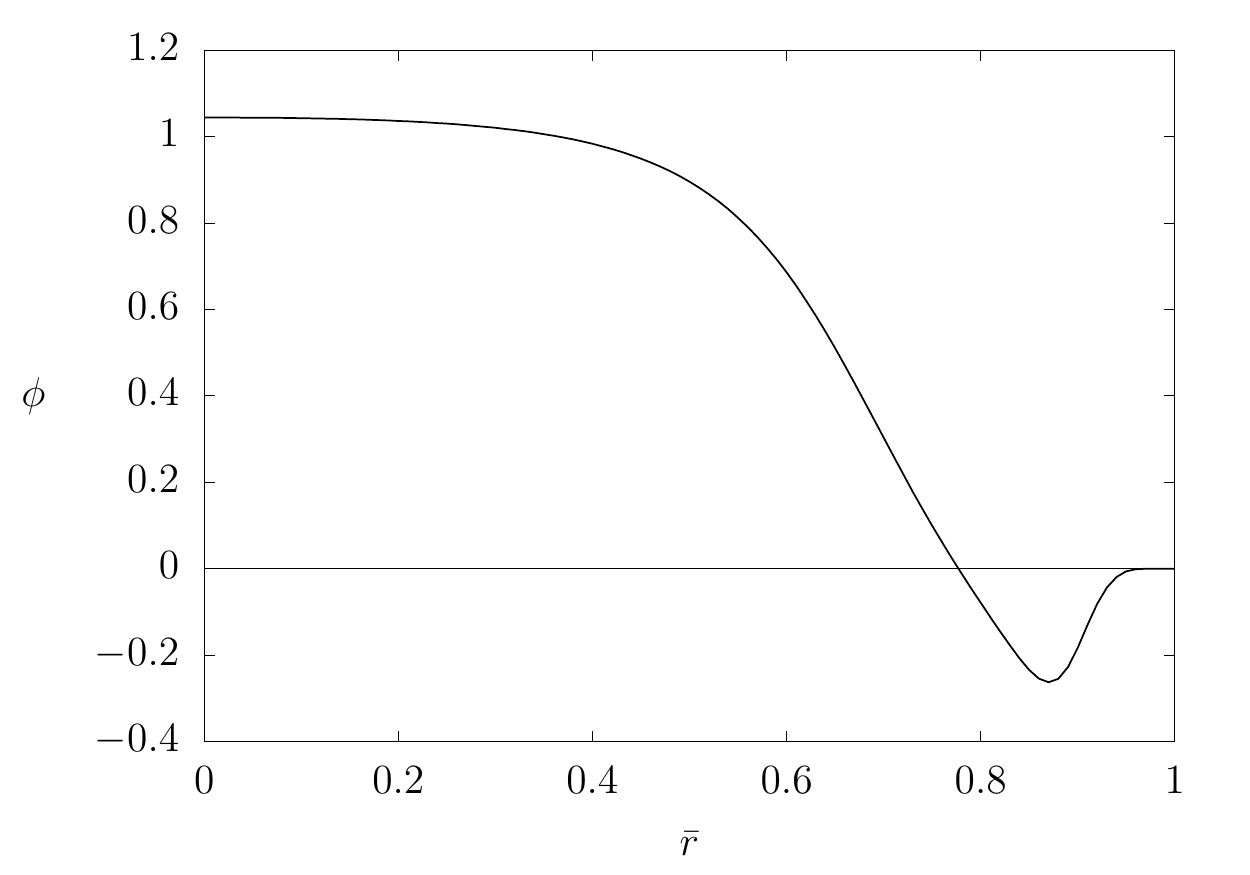}
%\caption{horse}
\label{nrqb}
\end{subfigure} \hspace{0.1\textwidth}
\begin{subfigure}{0.4\textwidth}
\includegraphics[scale=0.58]{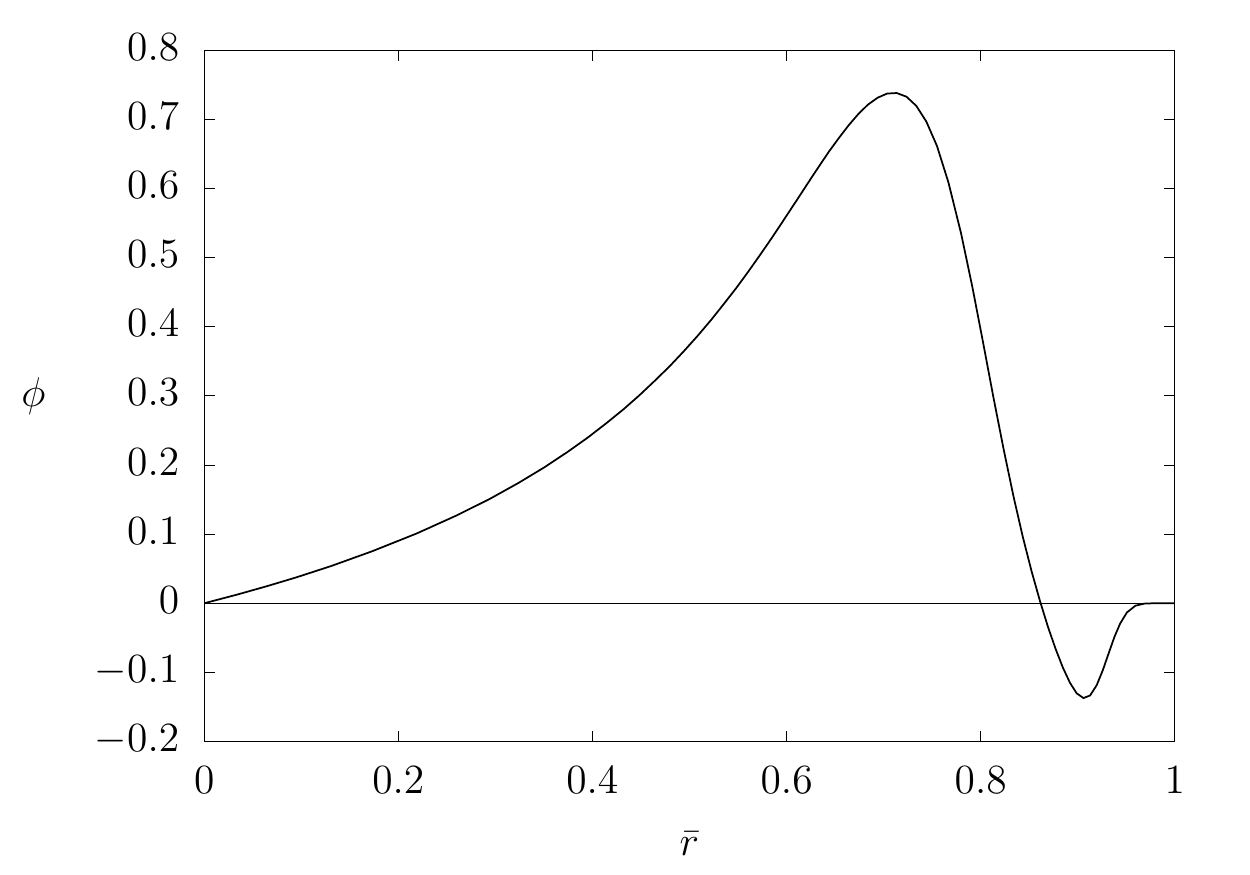}
%\caption{}
\label{rqb}
\end{subfigure}
\caption{\emph{Left:} Non-rotating radially excited Q-ball with boson
frequency $\omega_s=1.01$. 
\emph{Right:} Equatorial slice of a rotating radially excited Q-ball 
with boson frequency $\omega_s=1.01$}
\label{qb}
\end{figure}

We here investigate rotating boson star solutions 
with a first radial excitation. 
To obtain these we proceed in several steps.
Firstly a solution for the non-rotating Q-ball is found 
using the Runge-Kutta method of 4th order.
Then this solution is used to find the equivalent rotating Q-ball
as follows,
see Fig.~\ref{qb}. 
As stated above, the rotational quantum number $m$ must be an integer, 
but in order to get from $m=0$ to $m=1$ one can increase $m$ slowly,
starting from zero and passing through unphysical solutions
until $m=1$ is reached.
Once this value is attained we increase the coupling constant $\kappa$ 
from zero. This leads us from Q-balls to boson stars.  
Thus we obtain boson star solutions for different gravitational couplings.
All boson stars analyzed here have rotational quantum number $m=1$. 

It is worth to point out that it is difficult to construct
solutions with high accuracy at the turning points.
However, a good way to guarantee the trustability 
of a specific solution is to have the asymptotic calculations
of $M$ and $J$, eqs.~(\ref{asympmj}), equal the integral ones,
eqs.~(\ref{intm}) and (\ref{intj}).

\section{Solutions}
\label{s3}

We now discuss the physical properties of the rotating radially
excited boson star solutions obtained.
We start with their global charges, then address their
scalar field distributions, energy densities and pressures,
and end with some notes on their stability.

\subsection{Mass and Angular Momentum}

\begin{figure}[h!]
\begin{subfigure}{0.4\textwidth}
\includegraphics[scale=0.58]{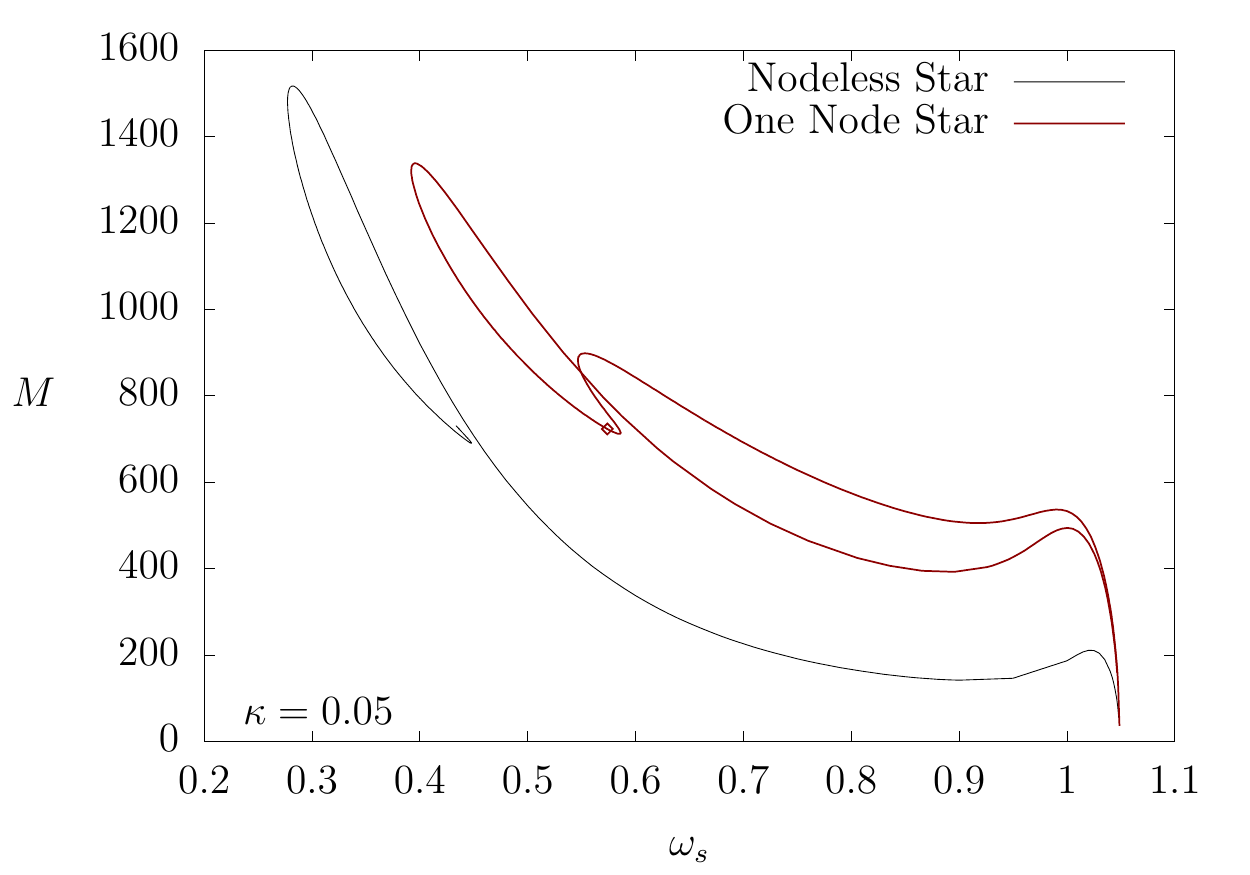}
\caption{}
\label{m0n1n}
\end{subfigure} \hspace{0.1\textwidth}
\begin{subfigure}{0.4\textwidth}
\includegraphics[scale=0.58]{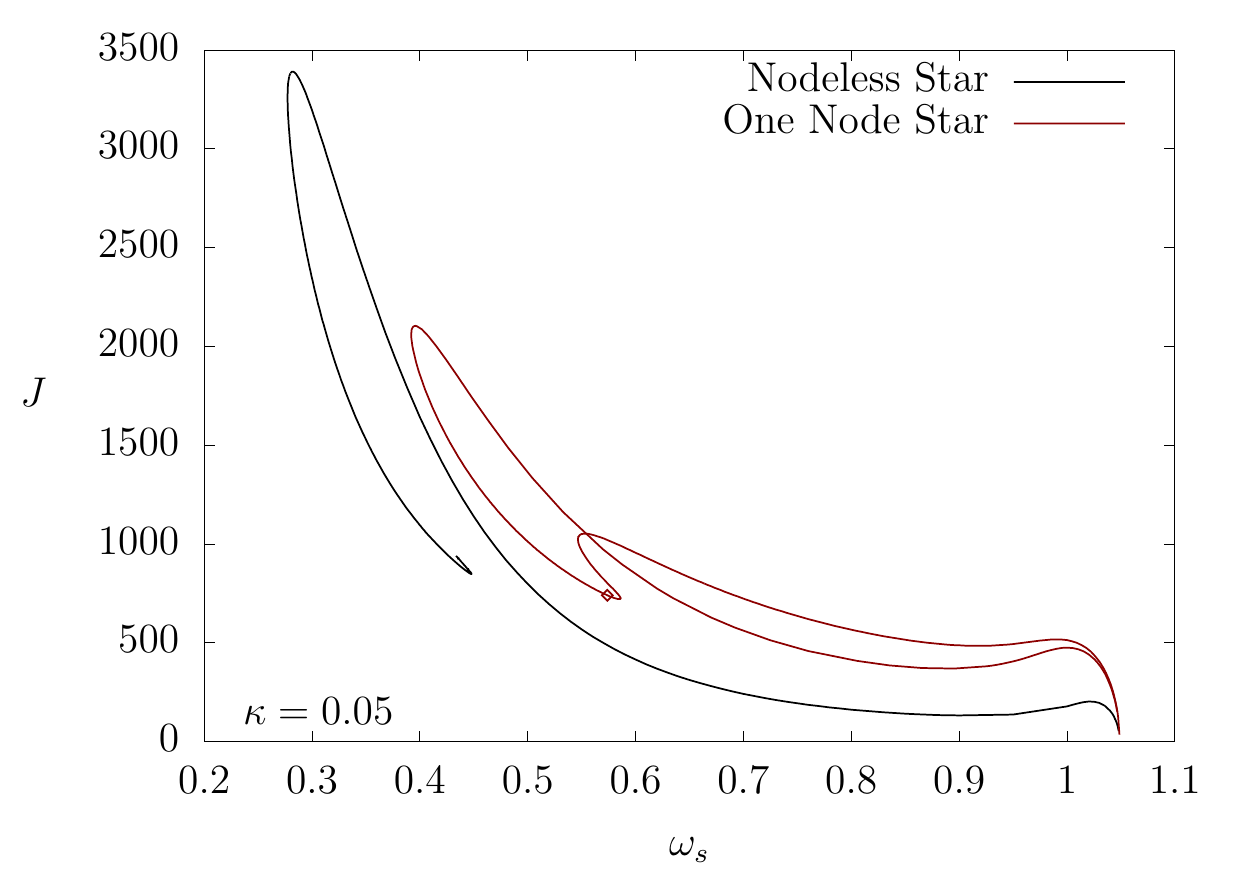}
\caption{}
\label{j0n1n}
\end{subfigure}
\caption{\emph{Left:} The mass $M$ of rotating boson stars ($m=1$)
with zero nodes (solid line) and with one node (dashed line)
versus the boson frequency $\omega_s$.
\emph{Right: } The angular momentum $J$ versus the boson frequency $\omega_s$
for the same set of solutions.
The diamond indicates the radially excited boson star solution 
with the maximal value of $\phi_0$.}
\label{omj0n1n}
\end{figure}

\begin{figure}[h!]
\begin{subfigure}{0.4\textwidth}
\includegraphics[scale=0.58]{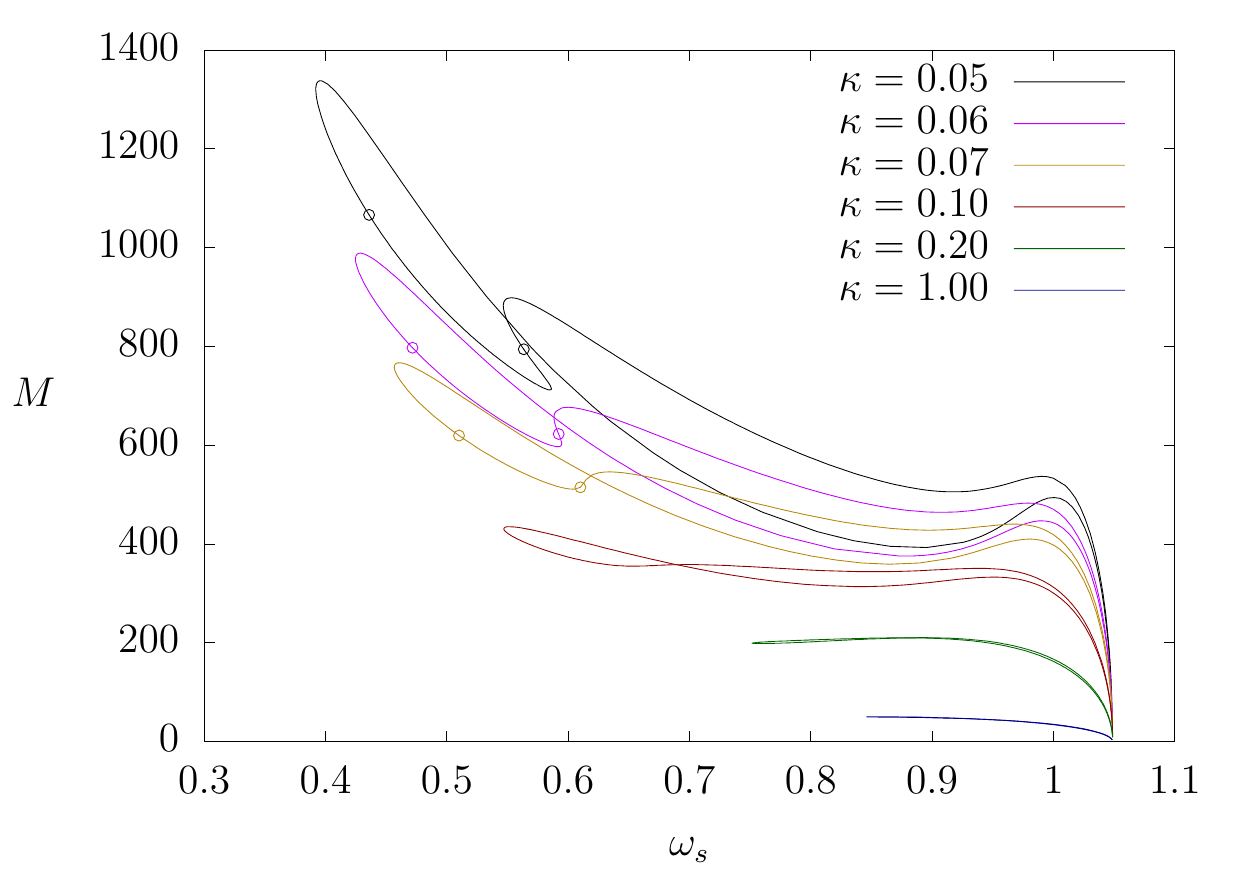}
\caption{}
\label{om}
\end{subfigure} \hspace{0.1\textwidth}
\begin{subfigure}{0.4\textwidth}
\includegraphics[scale=0.58]{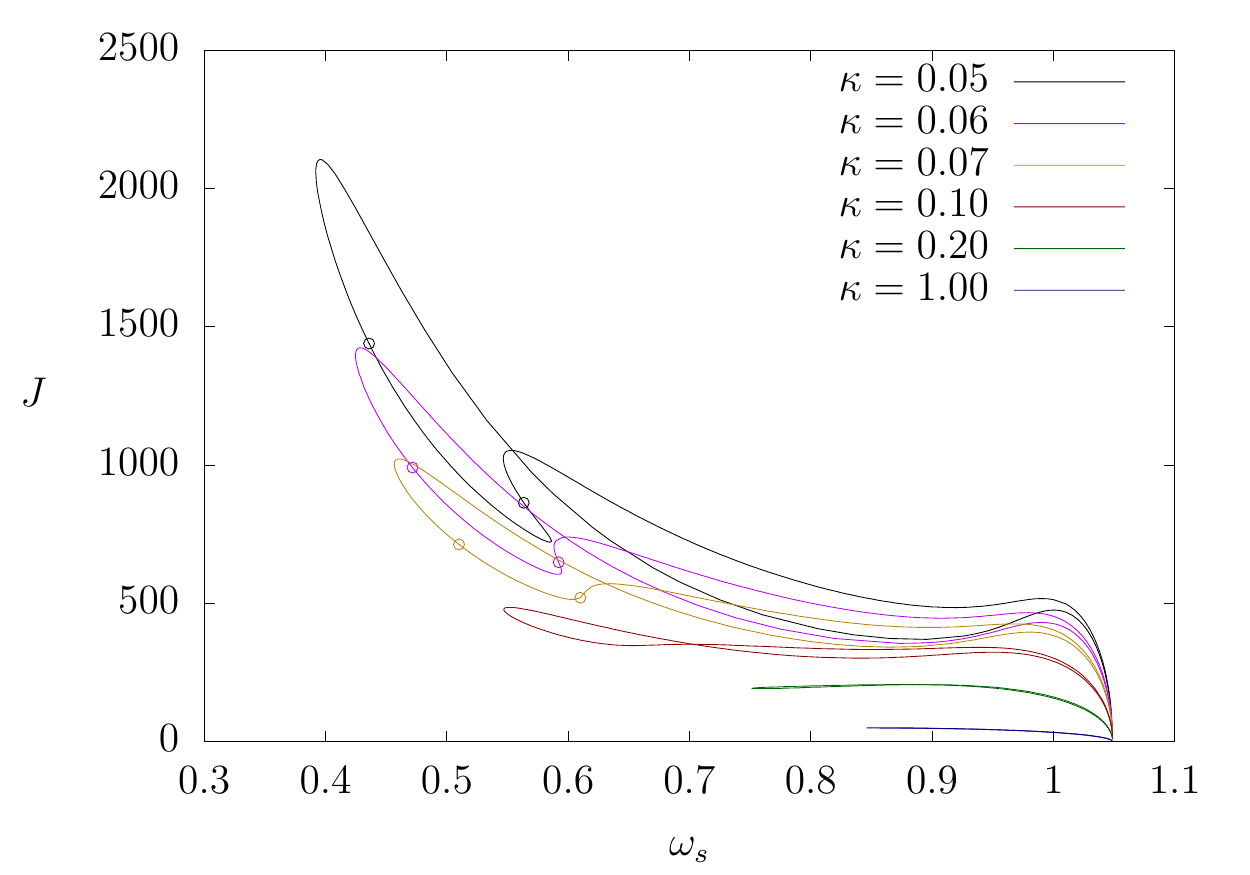}
\caption{}
\label{oj}
\end{subfigure}
\caption{\emph{Left:} The mass $M$ of the radially excited boson stars
versus the boson frequency $\omega_s$ 
for several values of the coupling $\kappa$. 
\emph{Right:} The angular momentum $J$ versus $\omega_s$ for 
the same set of solutions. 
The circles indicate the onset and termination of the
presence of ergoregions. 
}
\label{omj}
\end{figure}

To discuss the properties of rotating boson stars,
we exhibit in Figs.~\ref{omj0n1n} and \ref{omj} 
the mass $M$ and the angular momentum $J$
of several sets of boson stars
versus the boson frequency $\omega_s$
for fixed values of the coupling constant $\kappa$.
Let us recall, that the particle number $Q$ 
of rotating boson stars is
proportional to their angular momentum, $J=m Q$.
Non-rotating boson stars, 
both the fundamental solutions and the radially excited states, 
as well as fundamental rotating boson stars,
i.e., rotating boson stars without radial nodes, 
all present a spiraling behavior of the mass $M$ and 
the angular momentum $J$ versus the boson frequency $\omega_s$
\cite{LEE1992251,PhysRevD.72.064002}.
This spiraling behavior is seen in Fig.~\ref{omj0n1n} for
the set of fundamental $m=1$ boson star solutions with
coupling constant $\kappa=0.05$.

For this set of solutions the boson frequency cannot be used any more
to characterize the solutions uniquely. 
Nevertheless the central value of the field $\phi_0$
%(or its maximum value) 
can be used to characterize the non-rotating solutions
unless $\kappa$ is very small \cite{PhysRevD.85.024045}.
Likewise for fundamental rotating boson stars a good parameter choice 
is usually provided by the $m$-th derivative of the boson field 
$\phi_0=\partial_r^m\phi(0,\pi/2)/m!$, eq.~(\ref{phiorigin}),
which incorporates the leading behavior of the scalar field at the origin.

When besides rotation, which induces an angular excitation, 
further radial excitations are present, 
the spiraling behavior is no longer present in the set
of boson star solutions, as seen in Fig.~\ref{omj0n1n}. 
The curves for the mass and angular momentum of the 
radially excited rotating boson stars are only similar 
to those of the rotating boson stars without nodes up to the point,
where the onset of the spiral formation occurs for the latter. 
For the solutions with one radial node 
we notice, that when the spiral starts forming
it soon starts unfolding,
and we find at least two solutions for each boson frequency 
in the allowed domain. 
Thus the curves of the mass and angular momentum
versus the boson frequency form a non-trivial loop,
starting from and ending at the trivial solution at
$\omega_{\rm max}$.

For a given boson frequency, 
all solutions of a given set of radially excited
rotating boson star solutions
(specified by $m$ and $\kappa$),
including the intersection point within the loop, 
represent physically distinct solutions. 
For the same boson frequency and the same parameters, 
the radially excited boson stars 
possess a higher mass and angular momentum 
than the corresponding nodeless stars 
(for both branches of the radially excited solutions). 
Nevertheless, the set of solutions for the excited case 
is bounded from below by a higher value of $\omega_{\rm min}$ 
than the nodeless set. 
Furthermore, the latter presents a higher value 
for the maximum mass and angular momentum.

In order to find boson star solutions for bosons of different mass
the coupling constant $\kappa$ can be varied.
The lighter the individual boson is, 
the more massive will be the corresponding set of boson stars
\cite{LEE1992251}.
The greater the value of $\kappa$, i.e., the higher the boson mass, 
the greater is also the value of $\omega_{\rm min}$, 
for which bound solutions exist. 
Thus the domain of existence of boson star solutions
decreases with increasing coupling $\kappa$ 
as seen in Fig.~\ref{omj}.
This feature is observed generically in boson star models. 
We note, that for the three smaller values of $\kappa$ presented
in Fig.~\ref{omj} some of the solutions contain ergoregions.
The onset and termination of the presence of ergoregions there
is indicated by small circles.

Independent of the value of the coupling constant $\kappa$,
each set of excited rotating boson star solutions forms a loop. 
Only Q-balls, which are obtained in the limit of vanishing gravity,
are special in the sense, that for every value of $\omega_s$ 
there is at most one solution, 
i.e., one corresponding value of $Q$ and $M$. 
Therefore, for Q-balls the boson frequency can be used 
to uniquely parameterize the solutions,
which form a two parameter set of solutions $\{m,\omega_s\}$ 
once the behavior of the field is established, 
namely its radial and angular excitations.

\begin{figure}[h!]
\begin{subfigure}{0.4\textwidth}
\includegraphics[scale=0.58]{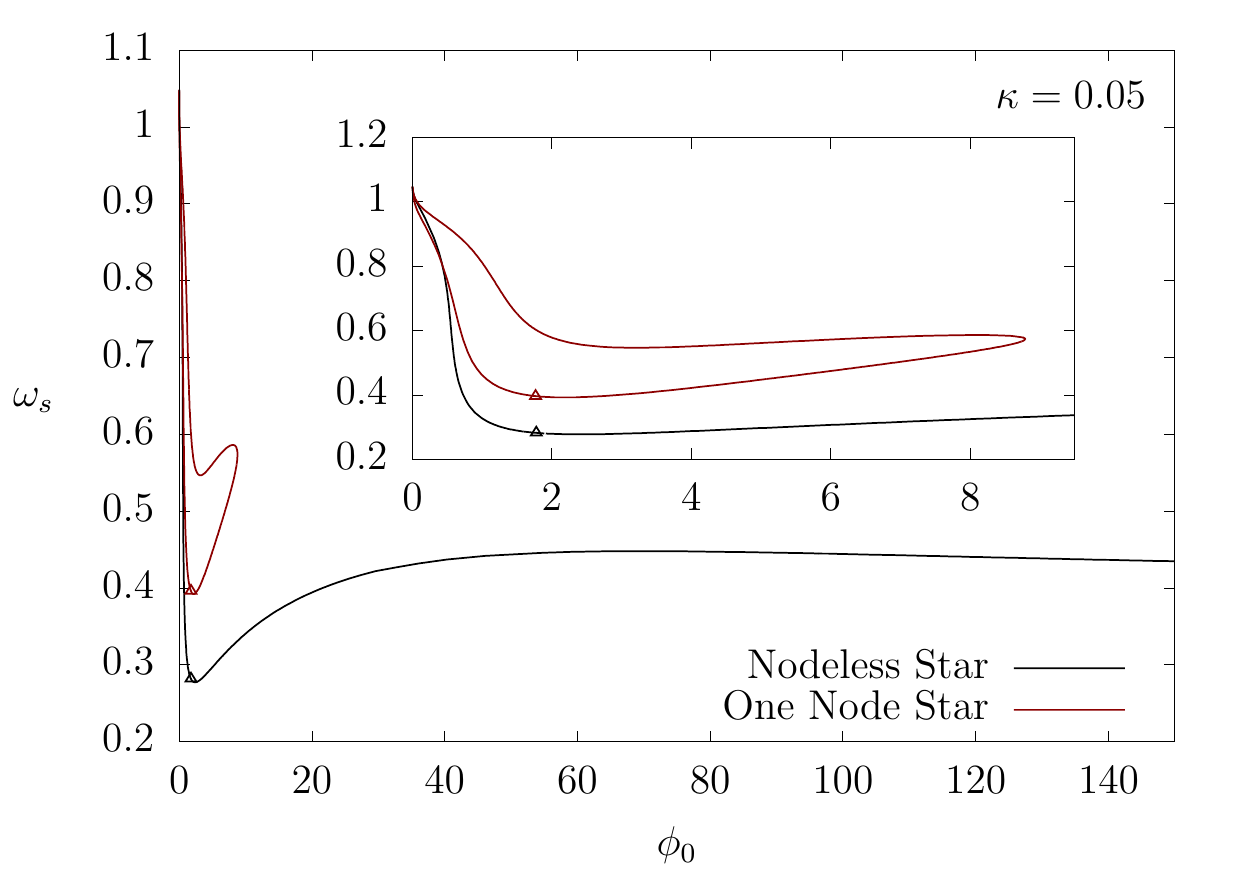}
\caption{}
\label{pron}
\end{subfigure}\hspace{0.1\textwidth}
\begin{subfigure}{0.4\textwidth}
\includegraphics[scale=0.58]{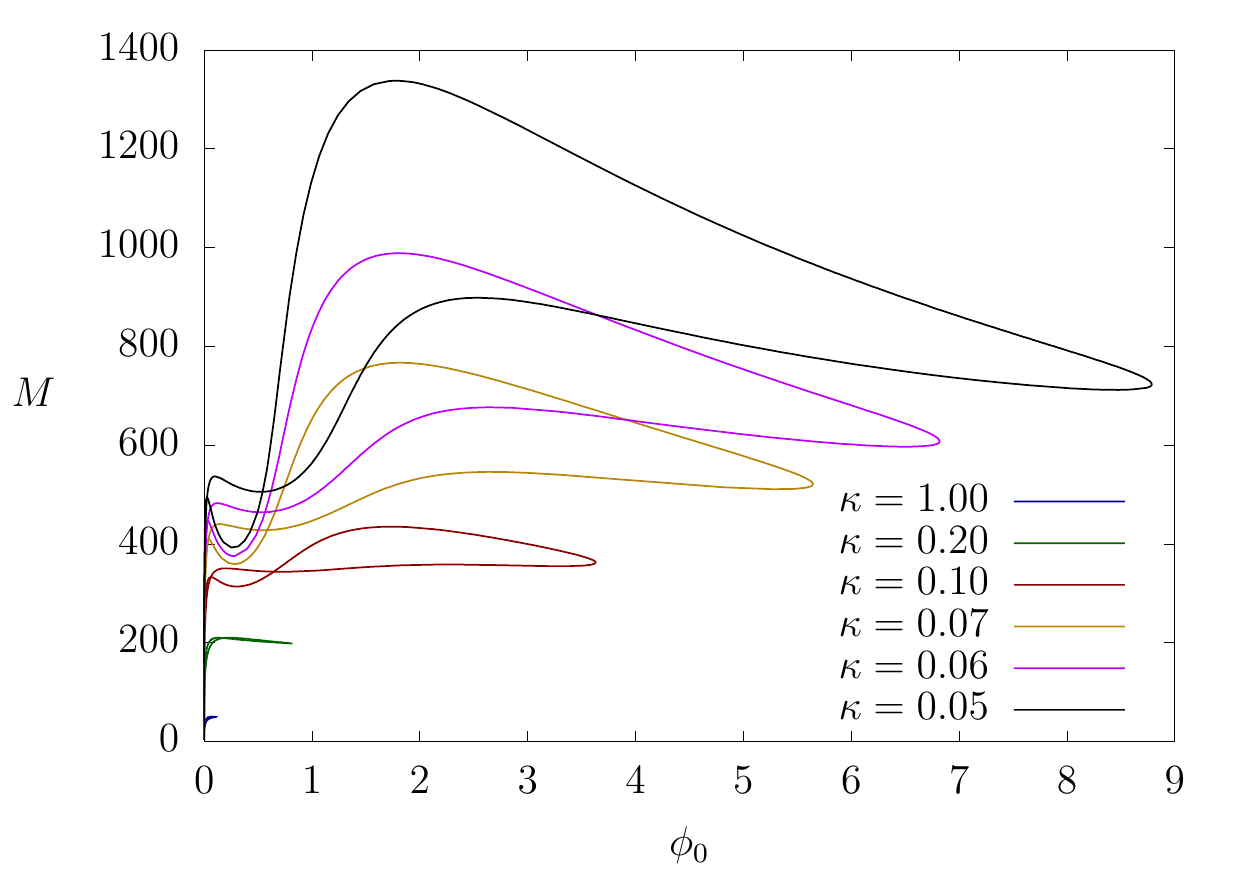}
\caption{}
\label{mphir}
\end{subfigure}
\caption{\emph{Left:} The boson frequency $\omega_s$ 
versus the value $\phi_0$, characterizing the boson field at the origin,
for the set of fundamental rotating boson star solutions
and those with one nodal excitation
for $\kappa=0.05$ and $m=1$.
\emph{Right:} The mass $M$ versus $\phi_0$
for the rotating radially excited boson stars for $m=1$
and several values of $\kappa$.
}
\label{pronmphir}
\end{figure}

As stated above
for spherically symmetric boson stars 
and rotating boson stars without nodes, 
the central value of the field $\phi_0$ 
is usually a good parameter 
to characterize the solutions uniquely.
In this case,
by increasing the value of $\phi_0$, 
first the spiral is approached and then
the solutions wind around the spiral
as $\phi_0$ keeps increasing.
In constrast, radially excited rotating boson stars
have an upper limit for $\phi_0$, 
beyond which $\phi_0$ decreases again towards zero.
This upper limit thus corresponds to a turning point 
in the $\omega_s$ versus $\phi_0$ diagram,
shown in Fig.~\ref{pron},
which also exhibits for comparison the dependence of
$\omega_s$ versus $\phi_0$
for the respective set of nodeless rotating boson stars.

Thus, in the case of rotating radially excited boson stars
for every value of $\phi_0$ there are two equilibrium configurations. 
This does not constitute an ill-posed problem, however,
because for each pair of solutions 
with the same value of $\phi_0$ for the boson field,
the boundary data of the metric functions are different. 
We also note that at the turning point, i.e., at the 
maximal value of $\phi_0$, there is only a single solution.
Likewise, the two branches converge to the trivial solution,
when the value of $\phi_0$ tends to zero
at the maximal value of the boson frequency.

This means that, although we can parameterize the solutions 
of such a set point by point in a non-trivial way, 
there is no physical quantity associated with the solutions 
that could be used as a parameter to characterize the
solutions uniquely.
We will therefore characterize the solutions by the two branches
of this diagram,
which we will simply refer to as upper and lower branch
(see section \ref{sub_stab}). 
The triangles in Fig.~\ref{pron} indicate the maximum mass solutions, 
and occur for the same value $\phi_0$ of the boson field.
Fig.~\ref{mphir} shows how the mass $M$ varies with $\phi_0$
for several values of the coupling $\kappa$.
Interestingly, when the coupling $\kappa$ is sufficiently small,
the maximum mass of a given set is found to reside at $\phi_0\sim 1.77$.
When the coupling is large, however,
the equilibrium configurations are limited 
by an upper bound of $\phi_0$ below that value.

\begin{figure}[h!]
\centering
\includegraphics[scale=0.58]{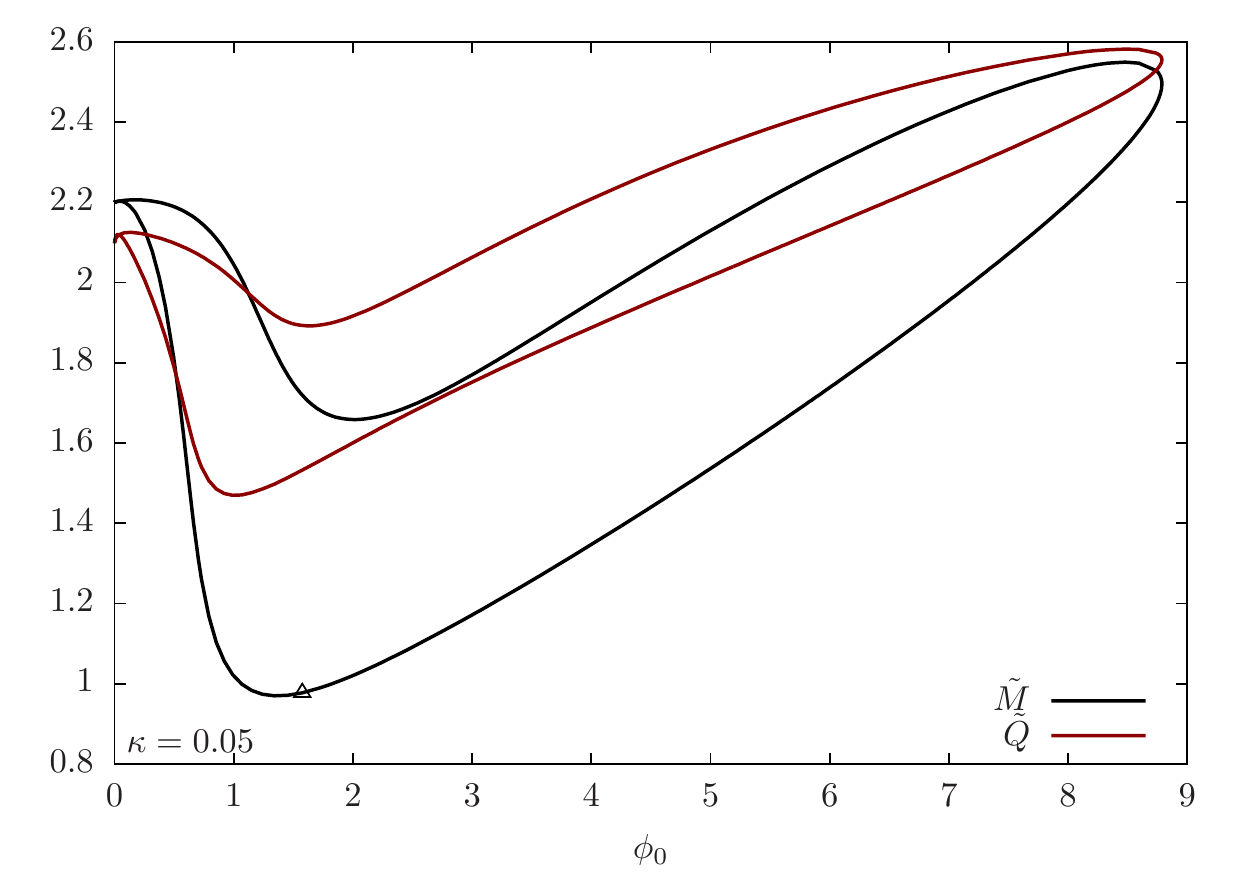}
\caption{Normalized mass $\tilde{M}$ and charge $\tilde{Q}$
versus the value $\phi_0$, characterizing the boson field at the origin,
for the set of rotating radially excited solutions
for $\kappa=0.05$ and $m=1$.
The triangle indicates the maximum mass solution.}
\label{normalmq}
\end{figure}

The normalization factor $N$, eq.~(\ref{normalfactor}),
can be used to normalize the 
mass and the angular momentum/charge (recall $J=Q$),
$\tilde{M}=M/N^2$ and
$\tilde{Q}=Q/N^2$,
shown in Fig.~\ref{normalmq}. 
The figure shows precisely two branches for each quantity, 
$\tilde{M}$ and $\tilde{Q}$,
which do not intersect themselves, in contrast to 
the branches of the mass $M$ shown in Fig.~\ref{pron}. 
Note, that the normalized charge attains its maximum
just to the left of the turning point.

\subsection{Boson Field Distribution and Charge Ratio}

\begin{figure}[h!]
%\centering
\begin{subfigure}{0.4\textwidth}
\includegraphics[scale=0.58]{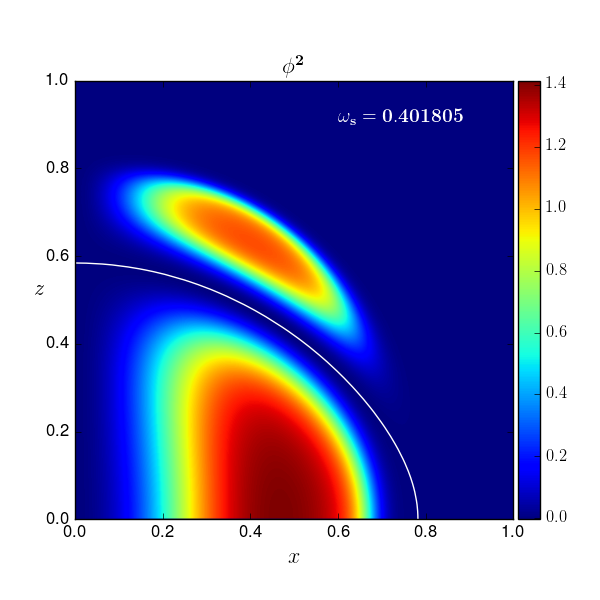}
\caption{Lower branch, $\phi_0=3.08316$}
\end{subfigure}  \hspace{0.1\textwidth}
\begin{subfigure}{0.4\textwidth}
\includegraphics[scale=0.58]{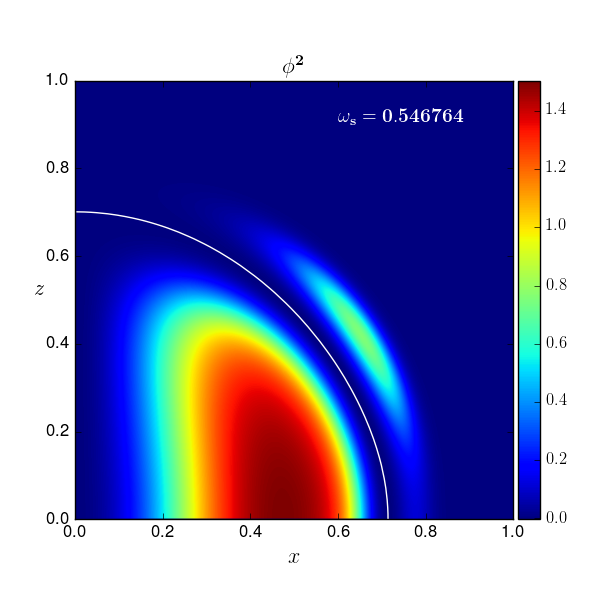}
\caption{Upper branch, $\phi_0=3.07778$}
\end{subfigure}
\\
\begin{subfigure}{0.4\textwidth}
\includegraphics[scale=0.58]{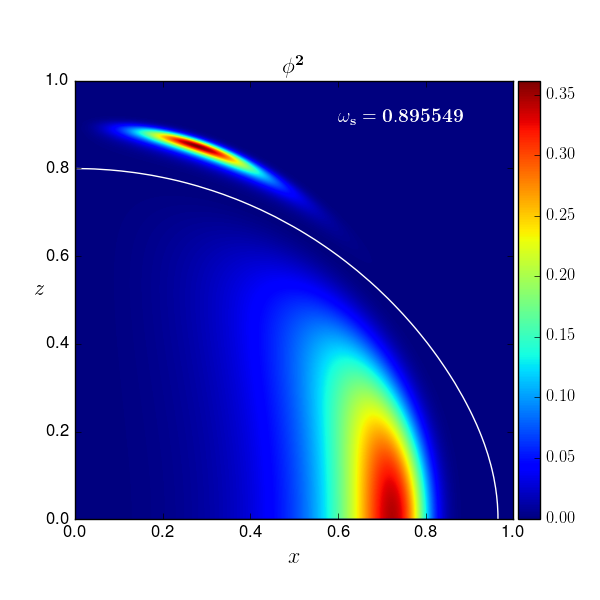}
\caption{Lower branch, $\phi_0=0.251513$}
\end{subfigure}   \hspace{0.1\textwidth}
% \hspace{0.2\textwidth}
\begin{subfigure}{0.4\textwidth}
\includegraphics[scale=0.58]{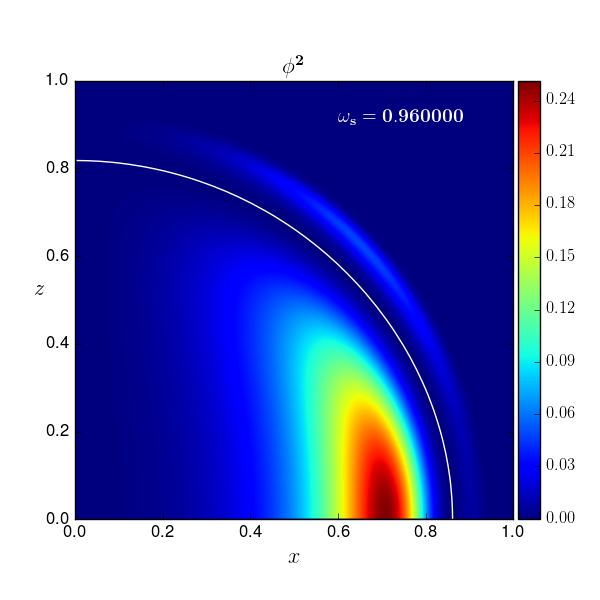}
\caption{Upper branch, $\phi_0=0.258202$}
\end{subfigure}
\caption{The distribution of the boson field
given by the square of its absolute value $\phi^2$ on the $xz$-plane,
where $z$ is the rotation axis and $x$ is an axis of the equatorial plane,
for rotating radially excited solutions (one node, $m=1$, $\kappa=0.05$).
The white curve represents the location of the node. 
\emph{Upper panels:} Solutions with similar values of $\phi_0$ near the turning point. 
\emph{Lower panels:} Solutions with similar values of $\phi_0$ for smaller $\phi_0$.}
\label{cmphi}
\end{figure}

Let us now focus on the set of rotating radially excited
solutions (one node, $m=1$) for coupling constant $\kappa=0.05$.
We then demonstrate, that boson star solutions 
with the same value of $\phi_0$
can nevertheless be very different. 
To this end we exhibit in Fig.~\ref{cmphi} 
the distribution of the boson field,
as given by the square of its absolute value $\Phi^*\Phi=\phi^2$, 
for two pairs of solutions with very similar boundary conditions. 
Thus the solutions of each pair correspond to solutions 
on the upper and lower branch,
located almost directly above each other 
in Figs.~\ref{pronmphir} and \ref{normalmq}.

\begin{figure}[h!]
\centering
\includegraphics[scale=0.58]{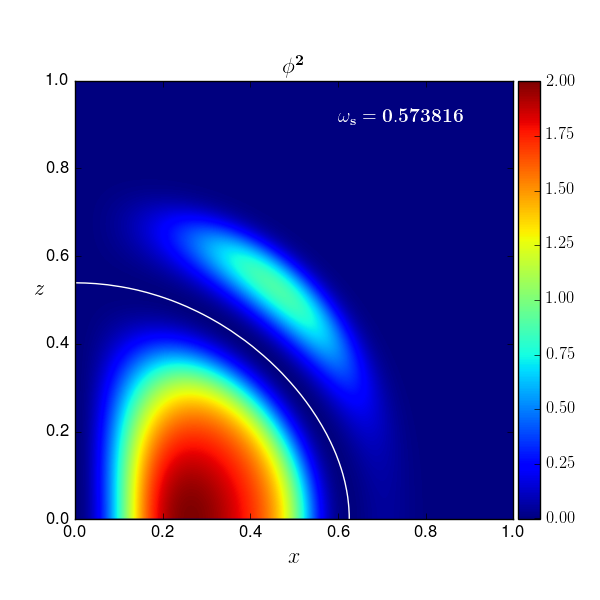}
\caption{The distribution of the boson field
given by the square of its absolute value $\phi^2$ on the $xz$-plane,
where $z$ is the rotation axis and $x$ is an axis of the equatorial plane,
at the turning point ($\phi_0=8.78668$)
for rotating radially excited solutions (one node, $m=1$, $\kappa=0.05$).
The white curve represents the location of the node.
}
\label{cmphim}
\end{figure}

A rotating radially excited boson star with one node 
can be thought of a system with two distinct shells. 
The inner shell always corresponds to values of $\phi>0$ 
and the outer one to values of $\phi<0$. 
Although the inner shell is always centered in the equatorial plane, 
the outer shell enjoys more freedom 
to center itself at different values of the polar angle $\theta$. 
For the solutions on the lower branch, 
the outer shell tends to have its maximum located
closer to the rotational axis, i.e., for higher values of $\theta$,
than for solutions on the upper branch.
Hence we note a higher eccentricity of the nodal curve,
represented in Fig.~\ref{cmphi} by a white line, 
for the solutions on the lower branch. 
The ratio between the negative minimum value of the field
and the positive maximum value of the field 
is overall higher on the upper branch and far from the turning point.
Fig.~\ref{cmphim} presents for the same set of boson stars
the solution at the turning point, where $\phi_0$ assumes
its maximum value. 
We note that for this solution
the nodal line is at a smaller distance from the origin
than for the two solutions close to it, exhibited in Fig.~\ref{cmphi}.

\begin{figure}[h!]
\centering
\includegraphics[scale=0.7]{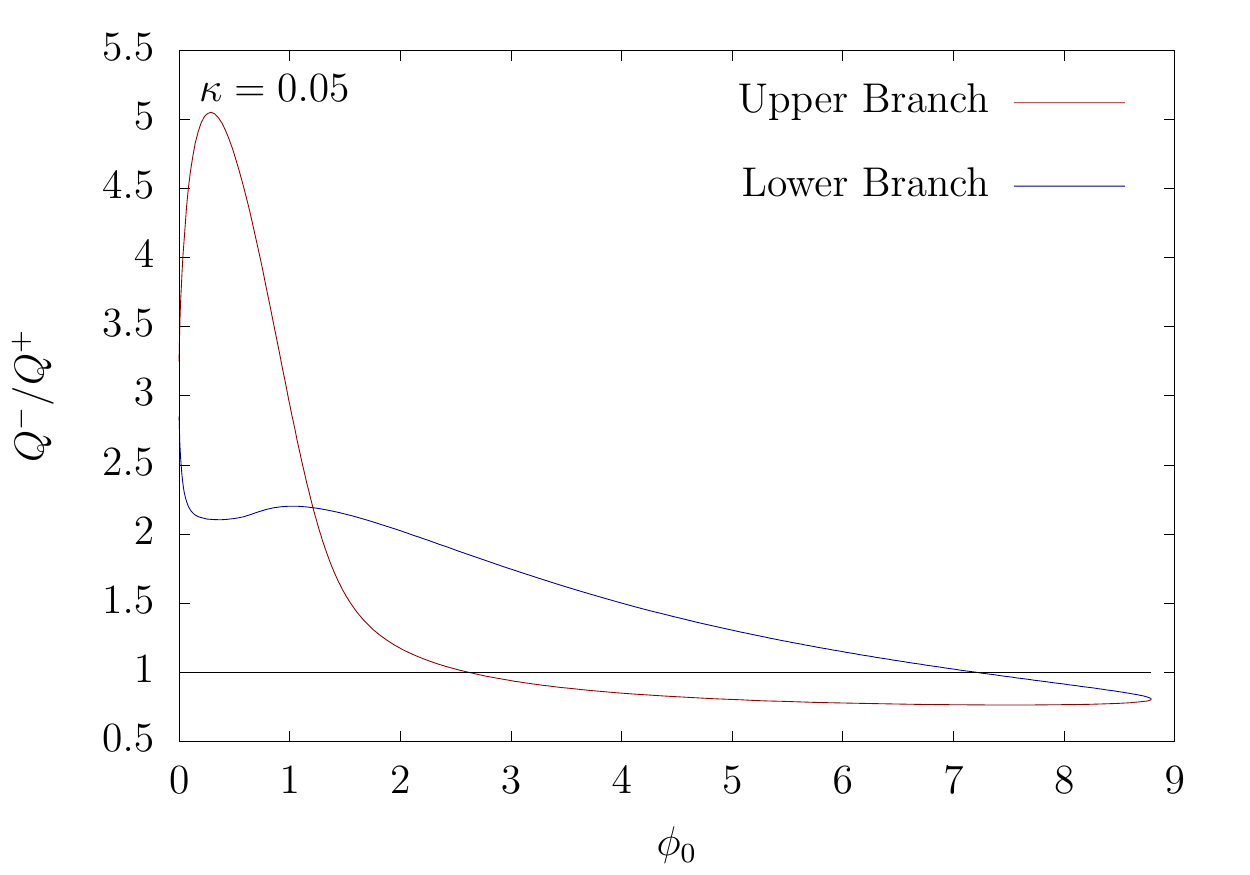}
\caption{Ratio $Q^-/Q^+$ of the number of particles
residing in the outer ($Q^-$) and inner ($Q^+$) shell
of rotating radially excited solutions (one node, $m=1$, $\kappa=0.05$).}
\label{ratioq}
\end{figure}

Let us now consider how the boson number (or charge) $Q$
is distributed among the two shells.
We therefore define $\phi^+$ to be the positive valued field 
comprehending the first shell, and $\phi^-$ the negative valued field 
comprehending the second shell. 
Then $Q^+=Q(\phi^+)$ is the charge accumulated in the first shell 
and $Q^-=Q(\phi^-)$ the charge located in the second one. 
In Fig.~\ref{ratioq} the ratio $Q^-/Q^+$ is shown. 
In most solutions the major part of the charge 
is concentrated in the outer shell, 
where in some cases it is as high as five times 
the charge residing in the inner shell. 
This feature changes in the vicinity of the turning point, 
where the configurations are characterized by a greater amount of charge 
in the inner shell. 
The observed distribution of the charges within the shells
might seem surprising at first,
when comparing with the field distribution,
but the field extends to infinity, and the metric determinant
in the volume integral gives more weight to the outer shell.

\subsection{Decomposition of the Field}

In order to better understand the radial excitations of the scalar field 
we will now decompose it into a complete set of functions 
of the coordinates. 
While the equations are not separable due to nonlinear effects, 
a decomposition into the set of associated Legendre functions
each coupled to a radial function can still be achieved. 
This decomposition for the boson function $\phi(r,\theta)$ is based on the
associated decomposition of the scalar field $\Phi$ itself, whose
spatial part would be decomposed into 
the spherical harmonics $Y_{\ell m}(\theta,\varphi)$ with $m=1$.
We therefore write
\[
\label{legendre}
\phi(r,\theta)=\sum_{\ell=|m|}^\infty R_\ell(r) 
P^m_\ell(\theta).\refstepcounter{equation}\tag{\theequation}
\]

Throughout this paper we have dealt only with the particular case of $m=1$. 
Clearly, the azimuthal quantum number $\ell$, 
which labels the field's angular excitations,
must then be greater or equal to one. 
Furthermore, the even parity character of the scalar field
results in $R_\ell=0$ for all even values of $\ell$. 
In the implementation of the decomposition 
the series was truncated at $\ell=11$, 
giving a very good approximation to the true solution 
with a maximal relative error of order $10^{-8}$ 
or better at any point.

Solutions for which the boson frequency $\omega_s$ is close to 
its maximum value possess a scalar field with very small amplitude.
The quartic and sextic terms in the potential (\ref{sixpot}) 
are then negligible across the whole domain. 
The Klein-Gordon equation eq.~(\ref{ekge}) then becomes linear. 
Furthermore, as these solutions approach the trivial one, 
the spacetime becomes almost flat. 

Still it is essential that there is a gravitational potential,
since in a flat spacetime, the linear Klein-Gordan equation
would not possess localized bound states.
%Under these circumstances a separation of variables could be done, 
%and represent a fairly good approximation of the field. 
In Minskowski spacetime, one would have a complete separation
of variables, and one could find solutions for which
%This approach would lead to solutions for which 
the bosons would all be in the same state, 
specified by the quantum numbers $\ell$ and $m$.
But these would correspond to free field configurations.
The analysis of the localized field configurations of the boson stars,
however, shows that the rotating radially excited
boson star solutions 
%do not contain such states.
are all represented by
%Instead we find that all boson fields represent
superpositions of states with several quantum numbers $\ell$
(while $m=1$).

In this linear regime, every term in the above decomposition 
represents a mode with quantum numbers $n, \ell, m$, 
%represents a pure state with quantum numbers $n, \ell, m$, 
where $n$ can be seen as to represent the principal quantum number.
In this case we recover the relationship $n=1+\ell+s$, 
where $s$ is the number of associated radial (spherical) nodes.

\begin{figure}[h!]
%\centering
\begin{subfigure}{0.4\textwidth}
\includegraphics[scale=0.58]{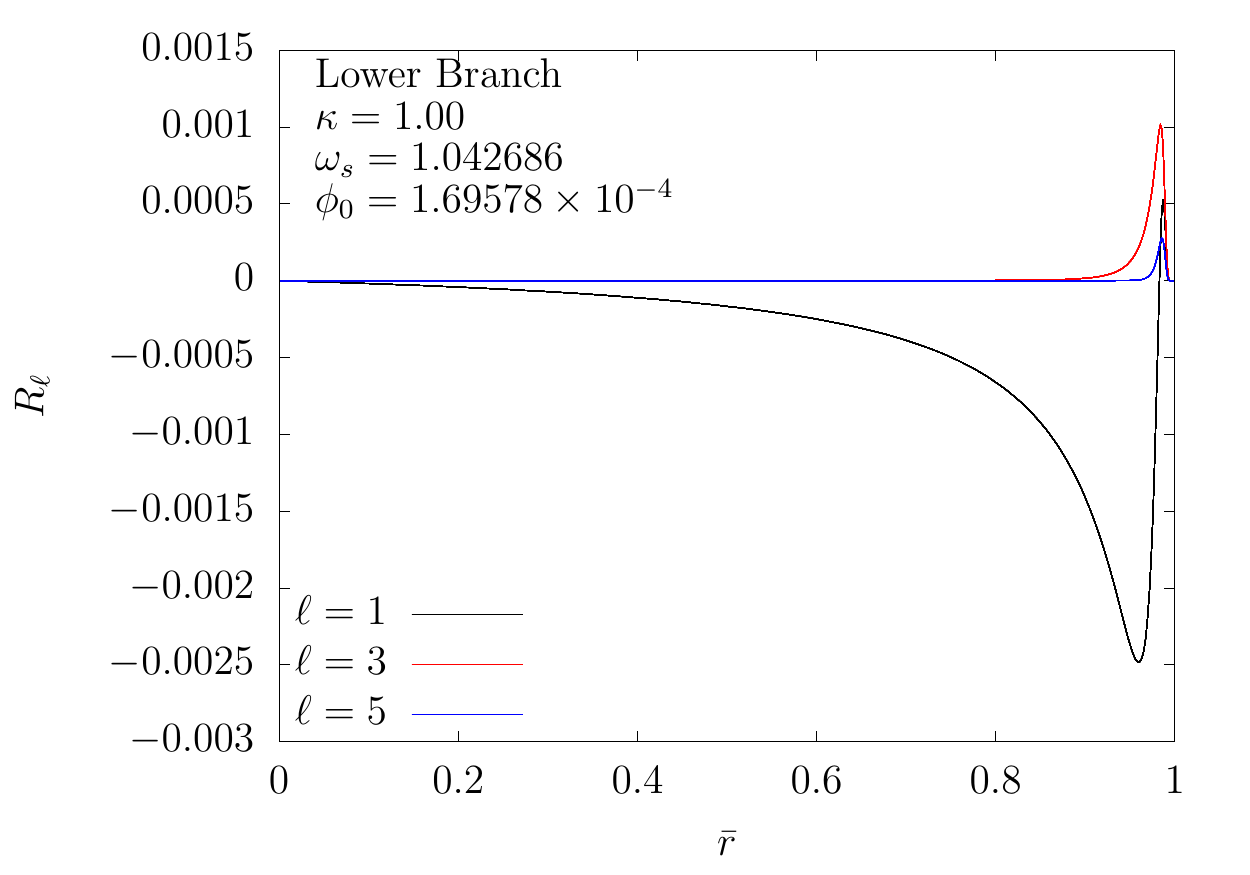}
\end{subfigure}  \hspace{0.1\textwidth}
\begin{subfigure}{0.4\textwidth}
\includegraphics[scale=0.58]{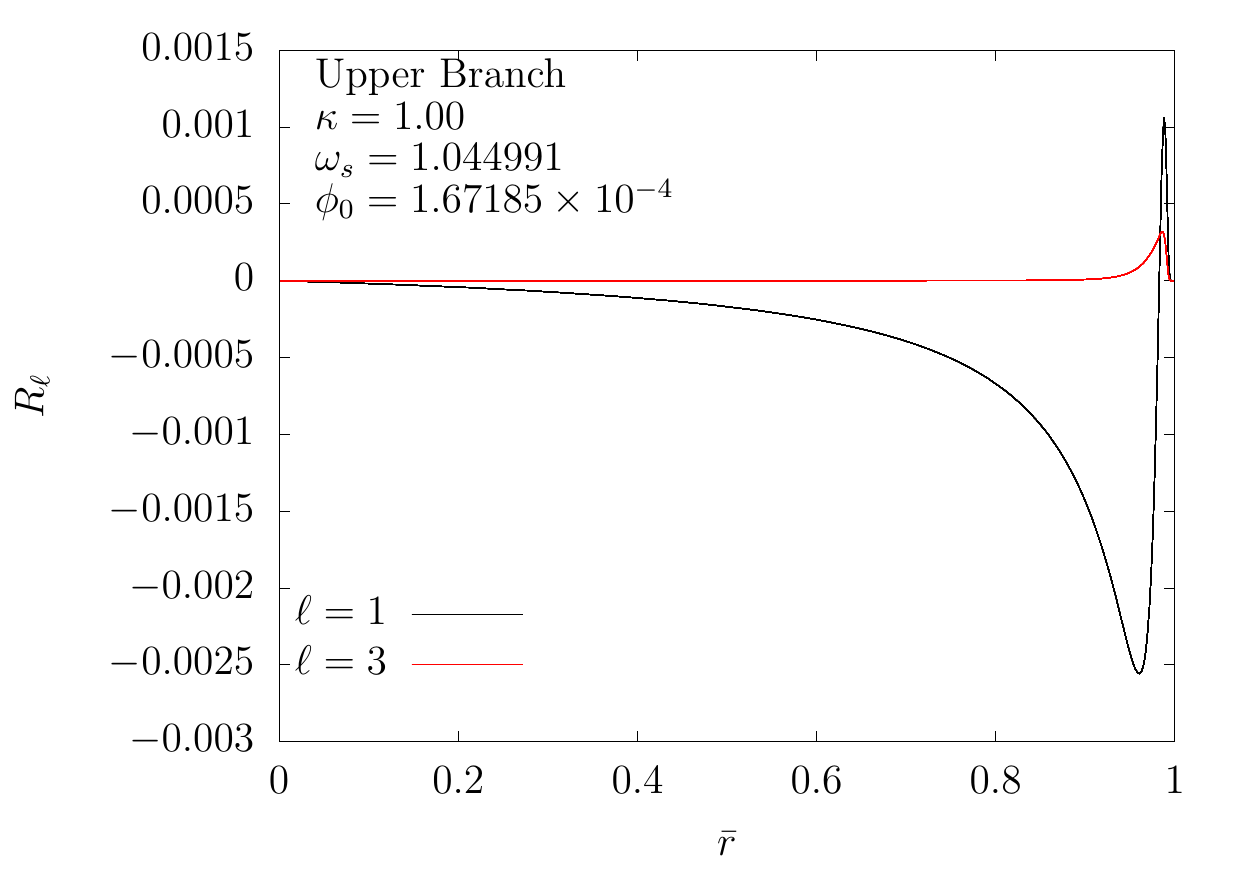}
\end{subfigure}
\caption{The lowest radial functions $R_{\ell}$ of the decomposition 
of the scalar field in terms of associated Legendre functions ($m=1$) 
versus the compactified radial coordinate $\bar r$
for rotating radially excited solutions
(one node, $\kappa=1.00$) in the linear regime
close to the maximum boson frequency.
}
\label{scdt}
\end{figure}

Fig.~\ref{scdt} shows two solutions in the linear regime
for coupling $\kappa=1.0$, and with similar values of the 
%boson frequency $\omega_s$ and the 
characteristic value of the field at the origin $\phi_0$,
belonging to different branches. 
Only the radial functions of the lowest $\ell$
modes are exhibited in the figure.
The others have very small amplitudes.
The lowest mode in the decomposition 
with respect to $\ell$ is given by $\ell=1, s=1$,
yielding $n=3$ in both cases.

The solution on the upper branch contains more particles 
in this mode than the solution on the lower branch, 
and fewer particles in higher $\ell$ modes.
For $\ell=7$ and $\ell=11$ the radial functions
on the lower and upper branches all contain one radial node. 
The remaining modes analyzed possess zero radial nodes. 
Thus, in the linear approximation 
valid for boson frequencies near the maximum value,
the solutions on the lower branch are composed of bosons 
in more excited (higher $\ell$) modes than those on the upper branch. 

\begin{figure}[h!]
%\centering
\begin{subfigure}{0.4\textwidth}
\includegraphics[scale=0.58]{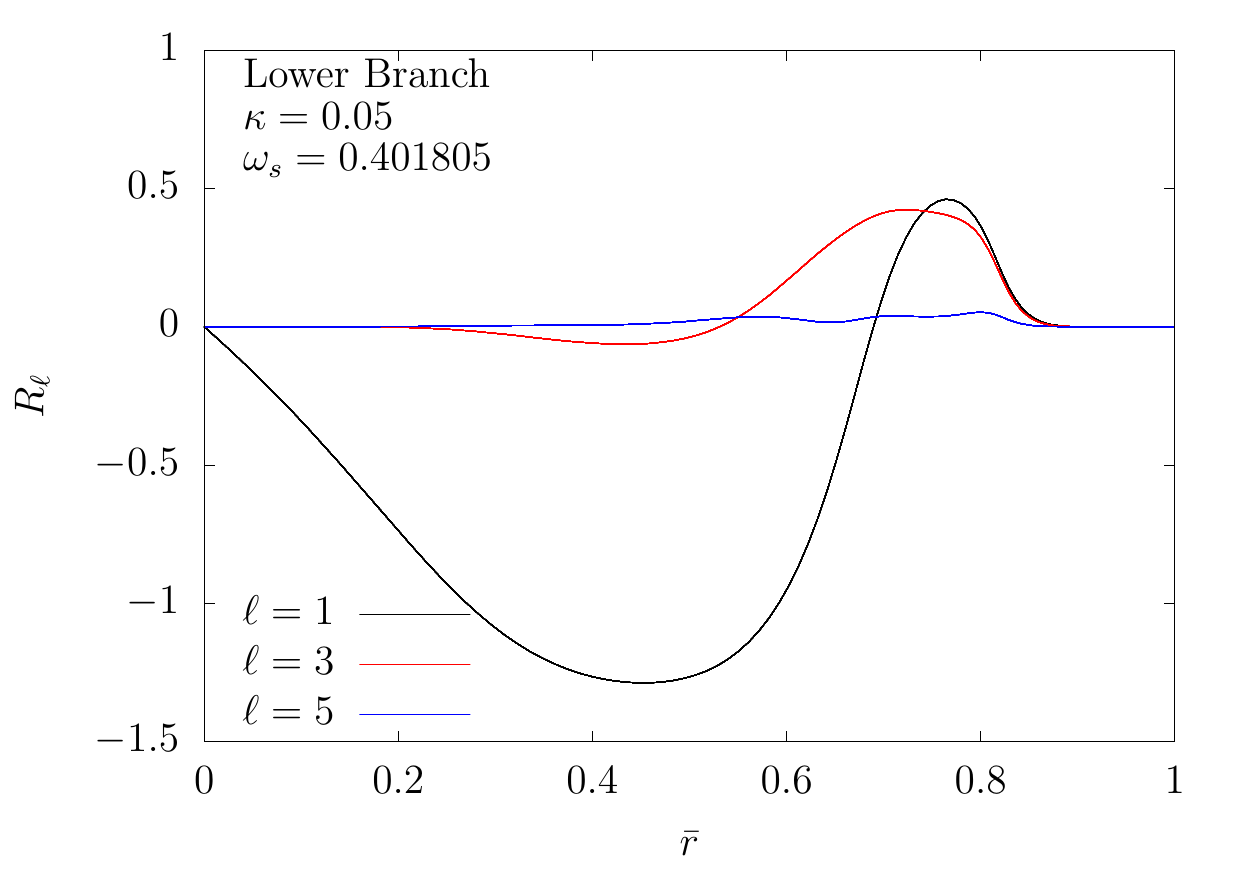}
\end{subfigure}  \hspace{0.1\textwidth}
\begin{subfigure}{0.4\textwidth}
\includegraphics[scale=0.58]{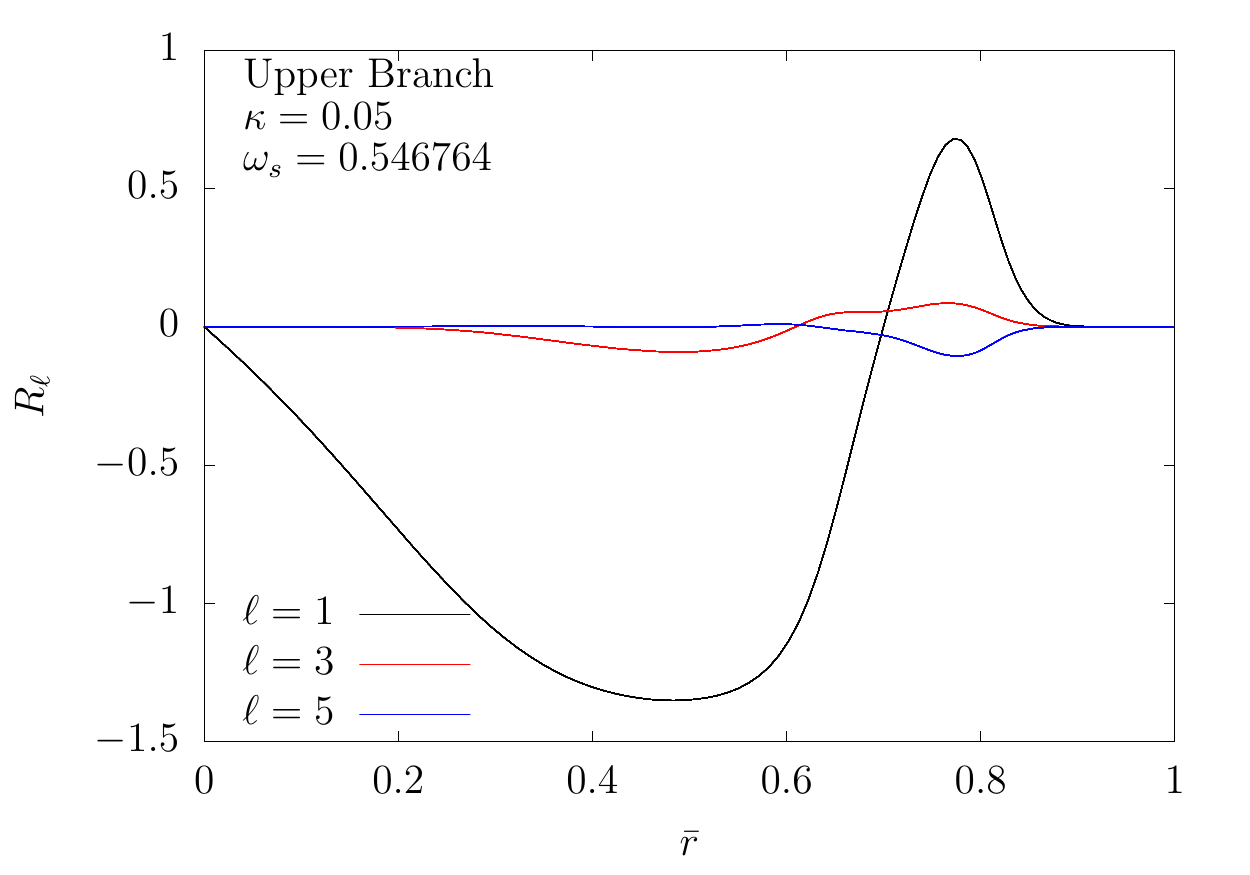}
\end{subfigure}
\\
\begin{subfigure}{0.4\textwidth}
\includegraphics[scale=0.58]{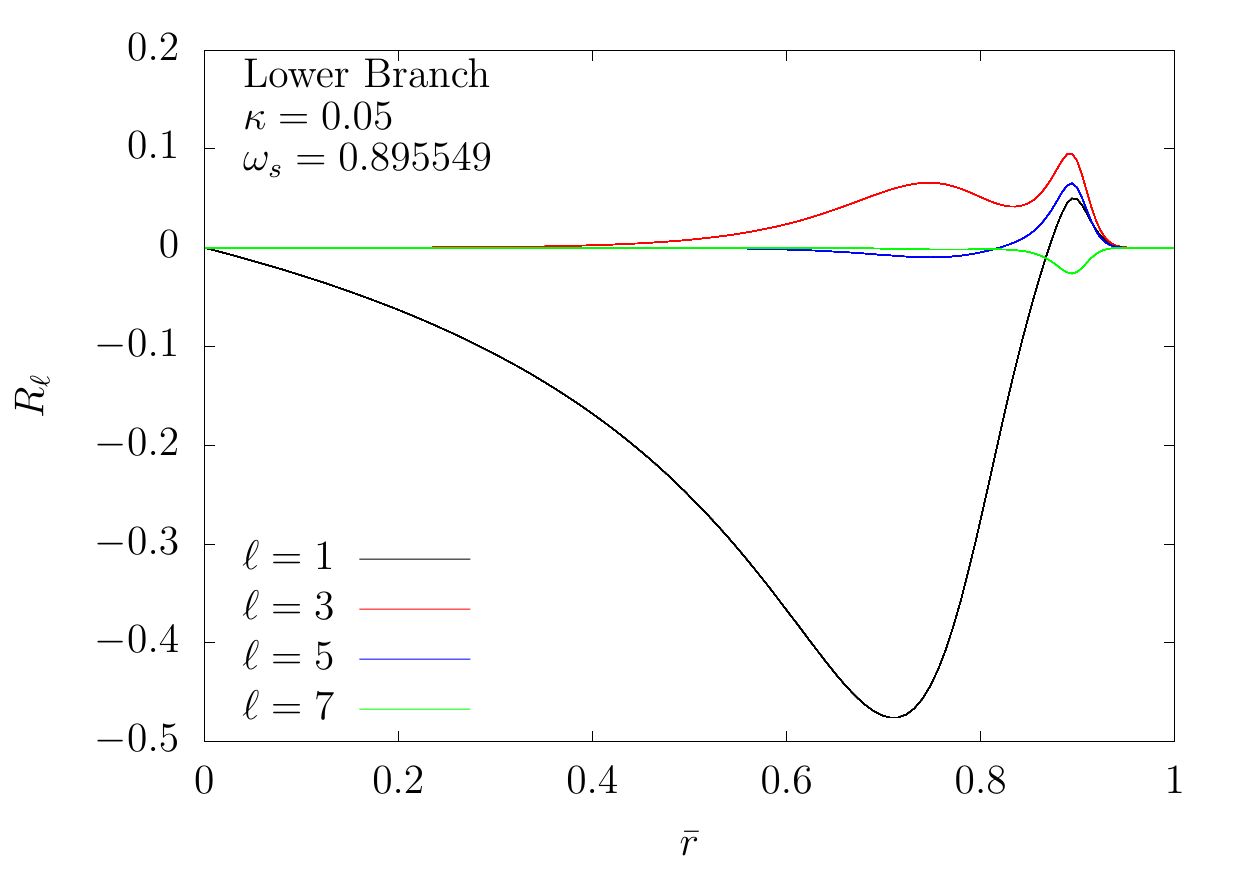}
\end{subfigure}   \hspace{0.1\textwidth}
% \hspace{0.2\textwidth}
\begin{subfigure}{0.4\textwidth}
\includegraphics[scale=0.58]{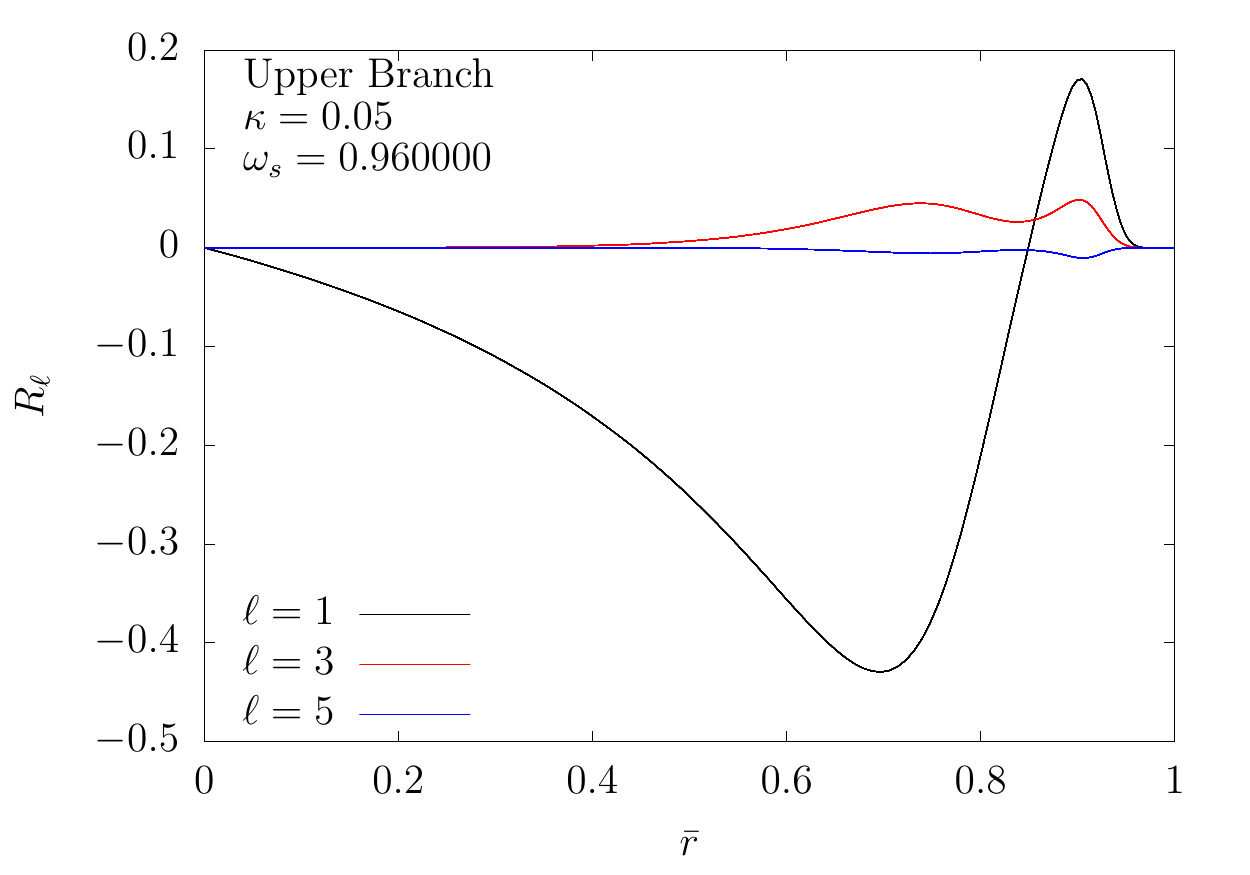}
\end{subfigure}
\caption{
The lowest radial functions $R_{\ell}$ of the decomposition
of the scalar field in terms of associated Legendre functions ($m=1$)
versus the compactified radial coordinate $\bar r$
for rotating radially excited solutions
(one node, $\kappa=0.05$).
\emph{Upper panels:} Solutions with similar value of $\phi_0$
near the turning point. 
\emph{Lower panels:} Solutions with similar value of $\phi_0$
for smaller $\phi_0$.
}
\label{scd}
\end{figure}

We now focus on the solutions analyzed in the previous subsection,
i.e., we address the solutions of the non-linear regime.
In Fig.~\ref{scd} the radial functions $R_{\ell}$ of the
decomposition, whose amplitude is not negligible,
are presented. 
We observe the same qualitative behavior 
that we saw for the solutions in the linear regime. 
%Nevertheless we can no longer interpret the individual terms 
%of the expansion (\ref{legendre}) as being the field of a pure state.
The larger contribution of the function $R_3$ 
for the solutions on the lower branch 
translates into the outer shell of the scalar field 
being closer to the rotational axis. 

\begin{figure}[h!]
\centering
\includegraphics[scale=0.58]{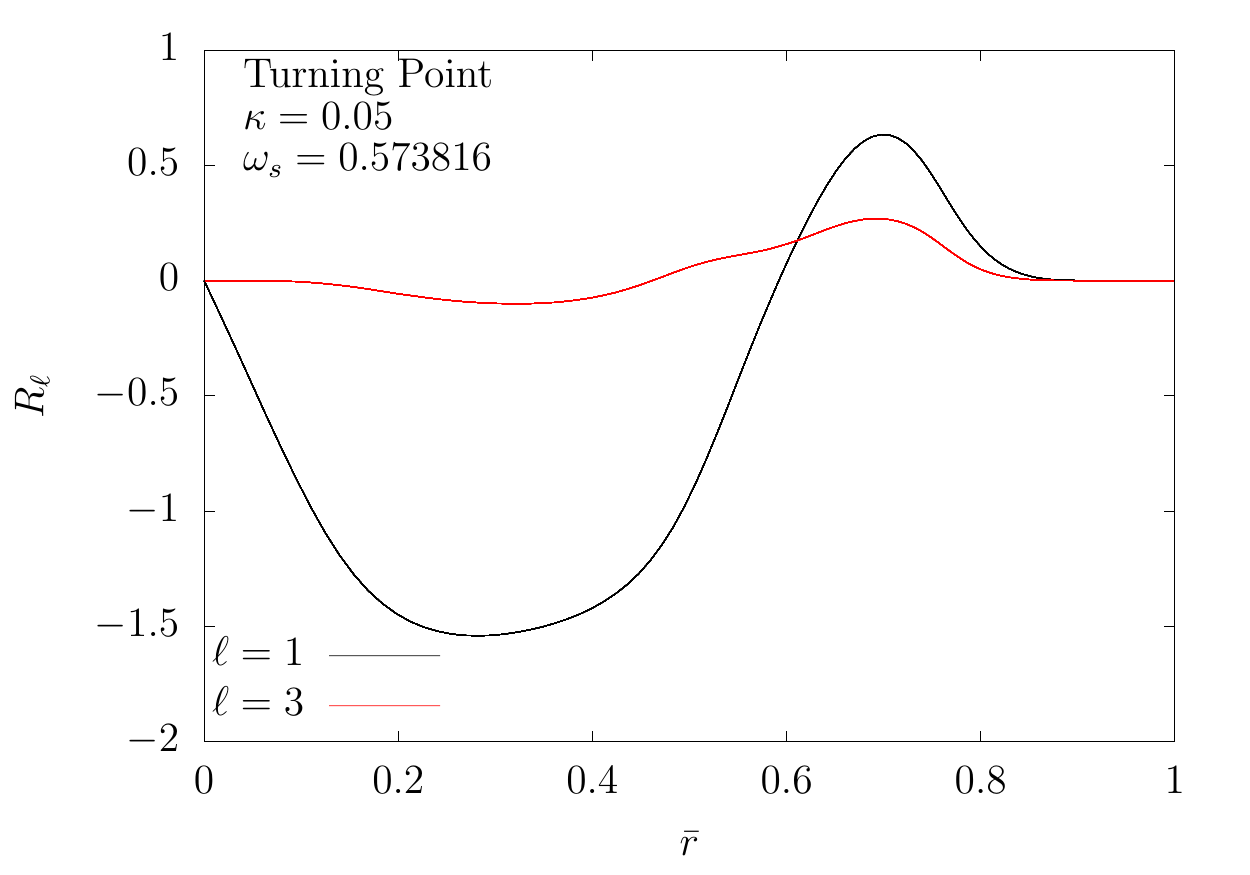}
\caption{
The lowest radial functions $R_{\ell}$ of the decomposition
of the scalar field in terms of associated Legendre functions ($m=1$)
versus the compactified radial coordinate $\bar r$
for the rotating radially excited solution at the turning point
(one node, $\kappa=0.05$).
}
\label{scdm}
\end{figure}

The above analysis reveals, that the solutions on
the upper and lower branches possess a different decomposition
into modes.
The decomposition for the solution at the turning point
is presented in Fig.~\ref{scdm}.
All modes higher than $\ell=3$ have very small amplitudes
in this case. 
We conclude with the observation 
that contrary to spherically symmetric boson stars, 
which are all represented by the lowest spherical harmonic $Y_{00}$,
(self interacting)
rotating boson stars cannot be formed by bosons in a single mode
$Y_{\ell m}$.

\subsection{Energy Density and Pressures}

In order to investigate the energy density and the pressures
of the rotating radially excited boson stars,
we take a \emph{vierbein} set $\hat{e}_a^\mu$, 
which are eigenvectors of energy momentum tensor. 
Then $u^\mu=\hat{e}_0^\mu$ is the four velocity 
of an element of the scalar fluid. 
The eigenvalues $\lambda_{(a)}$ correspond to the energy density 
and the pressures as measured by an observer comoving with the fluid, 
and are found through the relation
\begin{equation}
T_{\mu\nu}\hat{e}^\mu_a=\lambda_{(a)}\hat{e}_{\nu a},
\end{equation}
such that the diagonal energy momentum tensor 
takes the form $T^a_b=diag(\lambda_0,\lambda_1,\lambda_2,\lambda_3)$. 
The energy density is given by $\rho=-\lambda_0$, 
the radial pressure by $p_r=\lambda_1$,
and the tangential pressure by $p_\bot=\lambda_2=\lambda_3$,
since the scalar field constitutes an anisotropic fluid. 
The diagonal energy momentum tensor is then given by
\begin{align}
\label{tab}
T^a_b &=T_{\mu\nu}\hat{e}^{\mu a}\hat{e}^\nu_b \nonumber \\
      &=-\delta^a_b\left[\frac{1}{2}g^{\rho\tau}\left(\Phi^{\ast}_{, \rho}\Phi_{, \tau}+\Phi^{\ast}_{, \tau}\Phi_{, \rho}\right)+U\left(\lvert\Phi\rvert\right)\right]+\hat{e}^{\mu a}\hat{e}^\nu_b\left(\Phi^{\ast}_{, \mu}\Phi_{, \nu}+\Phi^{\ast}_{, \nu}\Phi_{, \mu}\right).
\end{align}

The equations of state for the radial and tangential pressure,
as derived from the above expression, read
\begin{equation}
p_r=\rho-2U, \qquad p_\bot=p_r-2\partial_\mu\phi\partial^\mu\phi,
\end{equation}
and take the same form as in the spherically symmetric case. 
At the nodal surface, 
all of these quantities have the same absolute value, 
i.e., $\rho=p_r=-p_\bot$. 
In contrast to the spherically symmetric boson stars, 
the tangential pressure is negative at the origin for 
rotating boson stars. 
The anisotropy factor $a_p$,
\begin{equation}
\label{anif}
a_p=\frac{p_r-p_\bot}{p_r},
\end{equation}
is not well behaved for either nodeless or 
radially excited rotating boson stars when $p_r=0$.
It always assumes 
the value $a_p=2$ at the nodal surface
(see the equations of state).

\begin{figure}[t!]
%\centering
\begin{subfigure}{0.4\textwidth}
\includegraphics[scale=0.58]{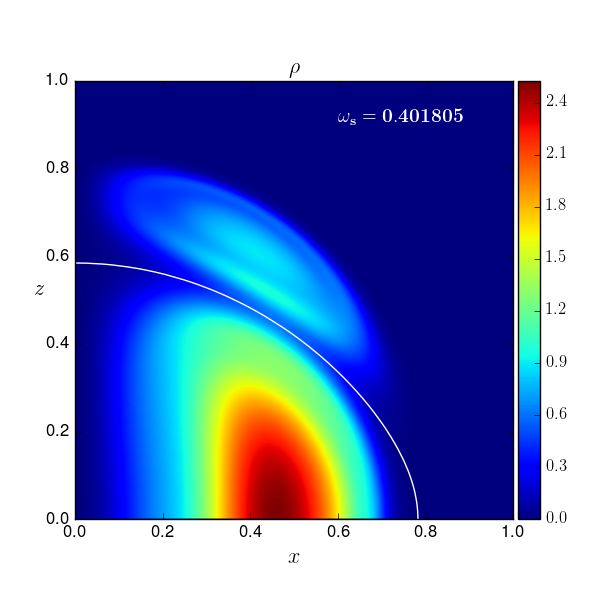}
%\caption{}
\end{subfigure}   \hspace{0.1\textwidth}
% \hspace{0.2\textwidth}
\begin{subfigure}{0.4\textwidth}
\includegraphics[scale=0.58]{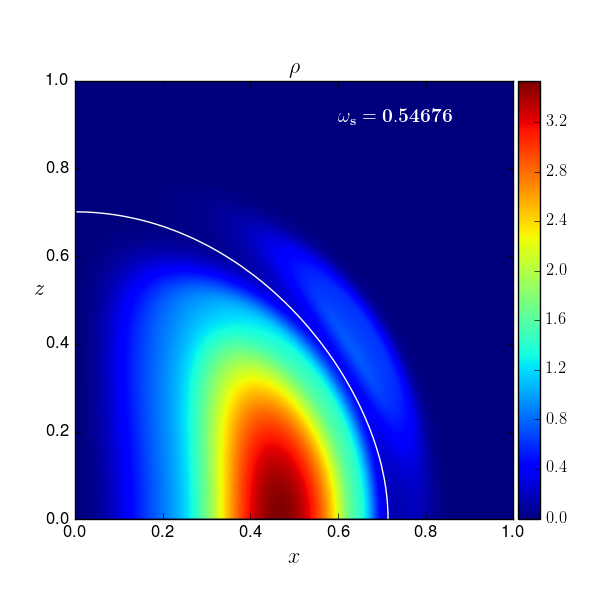}
%\caption{}
\end{subfigure}
\\
\begin{subfigure}{0.4\textwidth}
\includegraphics[scale=0.58]{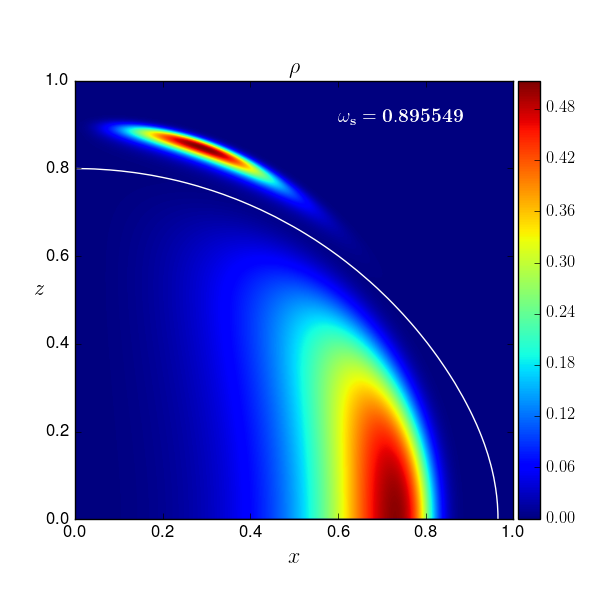}
%\caption{}
\end{subfigure}   \hspace{0.1\textwidth}
% \hspace{0.2\textwidth}
\begin{subfigure}{0.4\textwidth}
\includegraphics[scale=0.58]{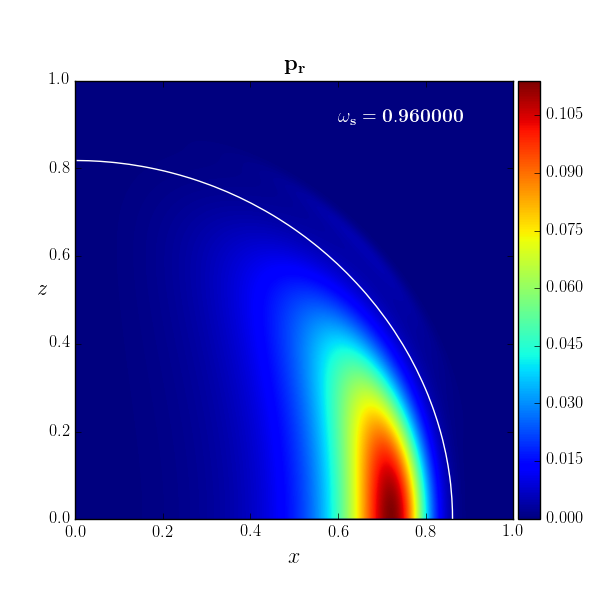}
%\caption{}
\end{subfigure}
\caption{Energy density $\rho$ 
as measured by an observer at rest with respect to the scalar fluid
on the $xz$-plane,
where $z$ is the rotation axis and $x$ is an axis of the equatorial plane,
for rotating radially excited solutions (one node, $m=1$, $\kappa=0.05$).
The white curve represents the location of the node.
\emph{Upper panels:} Solutions with similar values of $\phi_0$ near the turning point.
\emph{Lower panels:} Solutions with similar values of $\phi_0$ for smaller $\phi_0$.}
\label{cmrho}
\end{figure}

\begin{figure}[h!]
%\centering
\begin{subfigure}{0.4\textwidth}
\includegraphics[scale=0.58]{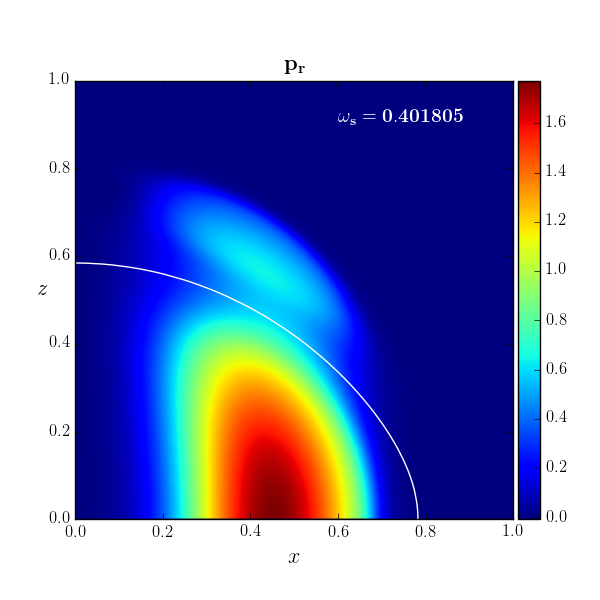}
%\caption{}
\end{subfigure}  \hspace{0.1\textwidth}
% \hspace{0.2\textwidth}
\begin{subfigure}{0.4\textwidth}
\includegraphics[scale=0.58]{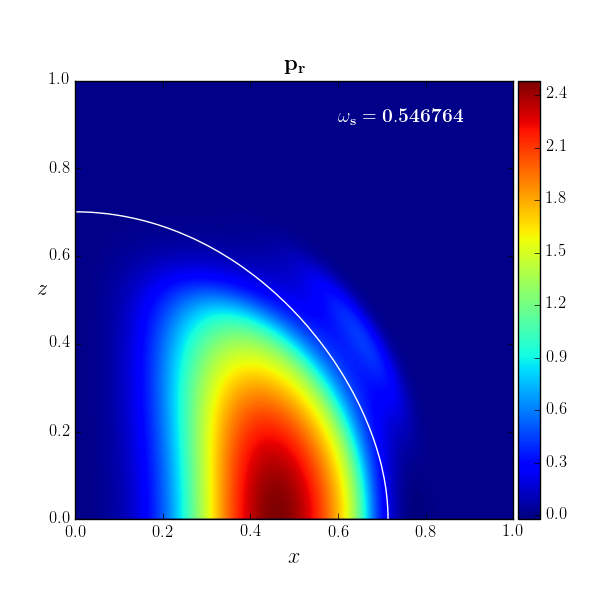}
%\caption{}
\end{subfigure}
\\
\begin{subfigure}{0.4\textwidth}
\includegraphics[scale=0.58]{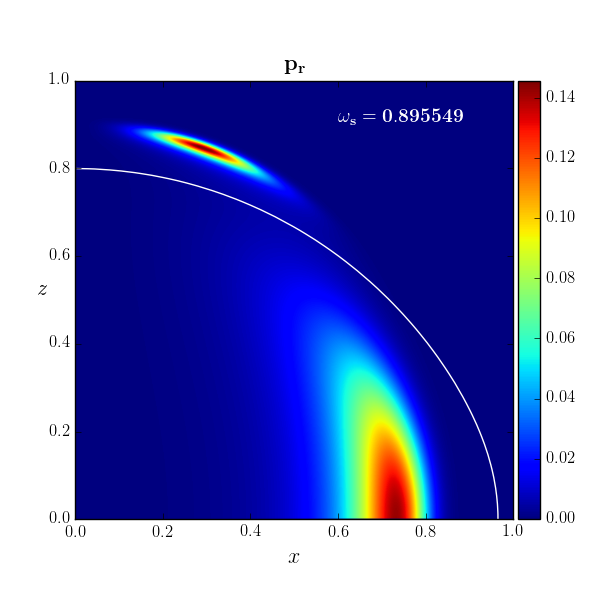}
%\caption{}
\end{subfigure}  \hspace{0.1\textwidth}
% \hspace{0.2\textwidth}
\begin{subfigure}{0.4\textwidth}
\includegraphics[scale=0.58]{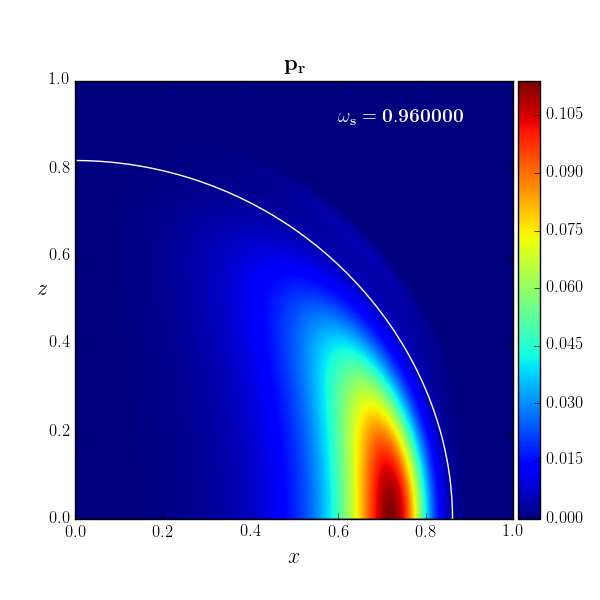}
%\caption{}
\end{subfigure}
\caption{
Radial pressure $p_r$
as measured by an observer at rest with respect to the scalar fluid
on the $xz$-plane,
where $z$ is the rotation axis and $x$ is an axis of the equatorial plane,
for rotating radially excited solutions (one node, $m=1$, $\kappa=0.05$).
The white curve represents the location of the node.
\emph{Upper panels:} Solutions with similar values of $\phi_0$ near the turning point.
\emph{Lower panels:} Solutions with similar values of $\phi_0$ for smaller $\phi_0$.}
\label{cmpr}
\end{figure}

\begin{figure}[h!]
%\centering
\begin{subfigure}{0.4\textwidth}
\includegraphics[scale=0.58]{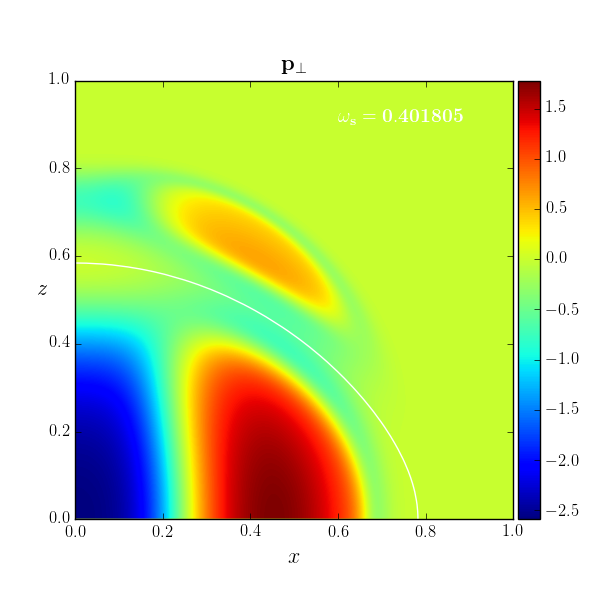}
%\caption{}
\end{subfigure}  \hspace{0.1\textwidth}
% \hspace{0.2\textwidth}
\begin{subfigure}{0.4\textwidth}
\includegraphics[scale=0.58]{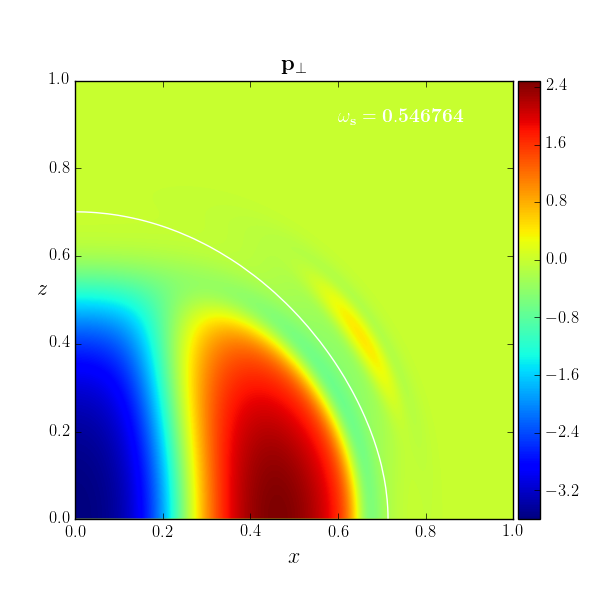}
%\caption{}
\end{subfigure}
\\
\begin{subfigure}{0.4\textwidth}
\includegraphics[scale=0.58]{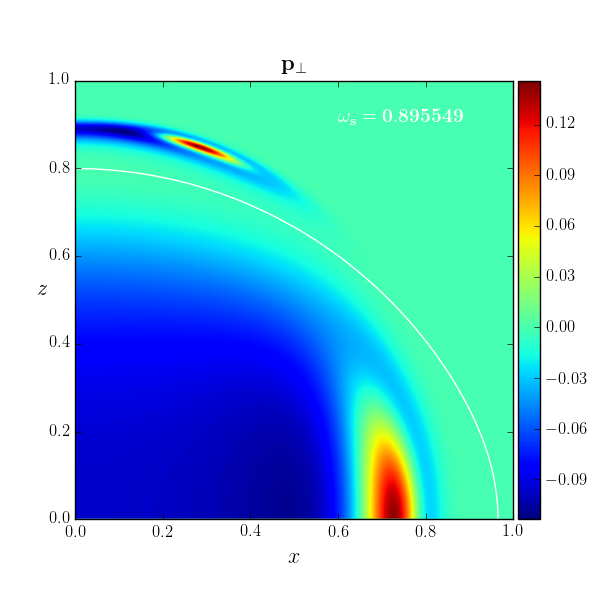}
%\caption{}
\end{subfigure}  \hspace{0.1\textwidth}
% \hspace{0.2\textwidth}
\begin{subfigure}{0.4\textwidth}
\includegraphics[scale=0.58]{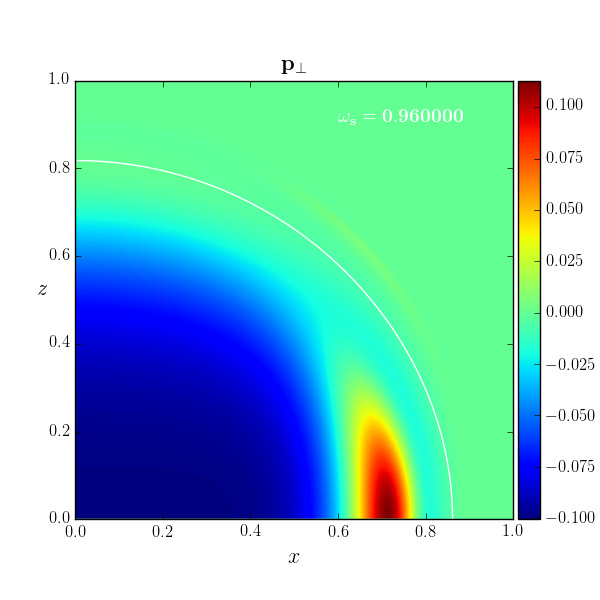}
%\caption{}
\end{subfigure}
\caption{
Tangential pressure $p_\bot$
as measured by an observer at rest with respect to the scalar fluid
on the $xz$-plane,
where $z$ is the rotation axis and $x$ is an axis of the equatorial plane,
for rotating radially excited solutions (one node, $m=1$, $\kappa=0.05$).
The white curve represents the location of the node.
\emph{Upper panels:} Solutions with similar values of $\phi_0$ near the turning point.
\emph{Lower panels:} Solutions with similar values of $\phi_0$ for smaller $\phi_0$.}
\label{cmpt}
\end{figure}

The general profile of the energy density and the pressures 
for the two pairs of solutions analyzed before 
are shown in Figs.~\ref{cmrho}, \ref{cmpr} and \ref{cmpt}. 
The energy density and the radial pressure are distributed 
in a similar manner as the boson field squared.
Likewise, both are zero at the origin. 
The tangential pressure, however, 
starts with negative values close to the origin 
and turns positive only in the vicinity of the center
of the inner and outer shell, respectively.

\begin{figure}[h!]
\begin{subfigure}{0.4\textwidth}
\includegraphics[scale=0.58]{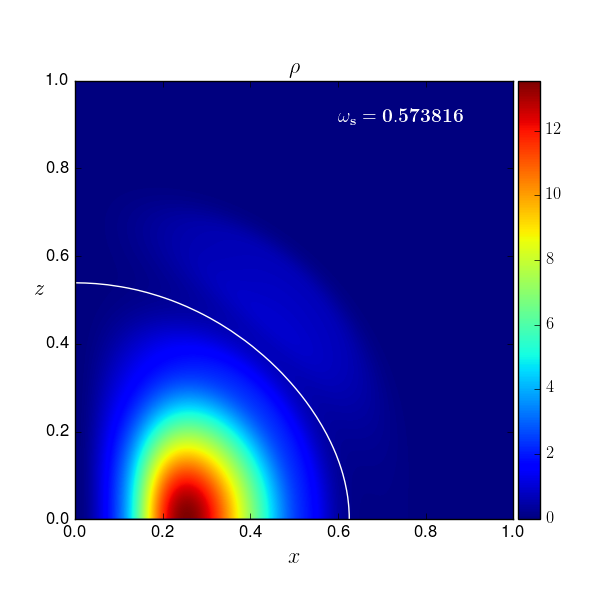}
%\caption{Energy density of the most dense boson star.}
\end{subfigure}
\hspace{0.1\textwidth}
\begin{subfigure}{0.4\textwidth}
\centering
\includegraphics[scale=0.58]{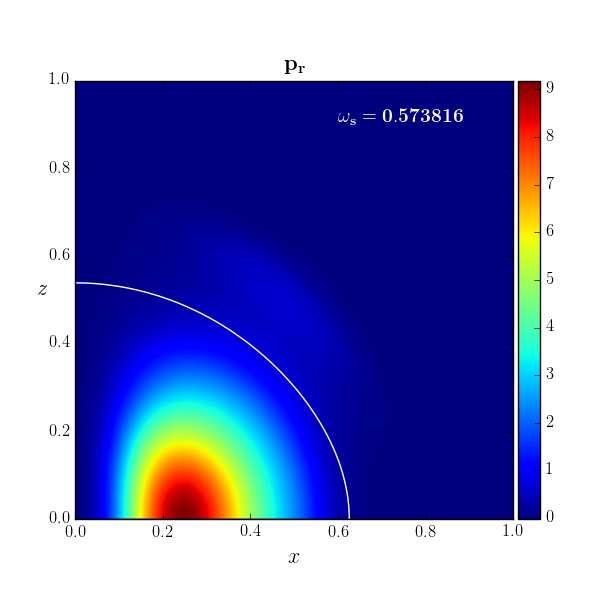}
%\caption{Radial pressure of the most dense boson star.}
%\label{prm}
\end{subfigure}
\begin{subfigure}{0.4\textwidth}
\centering
\includegraphics[scale=0.58]{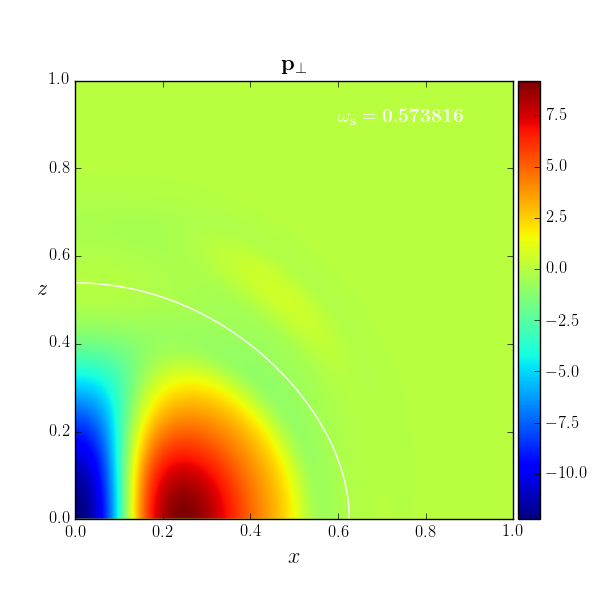}
%\caption{Tangential pressure of the most dense boson star.}
%\label{ptm}
\end{subfigure}
\caption{
Energy density $\rho$ (\emph{upper left panel}),
radial pressure $p_r$ (\emph{upper right panel}),
and tangential pressure $p_\bot$ (\emph{lower panel})
as measured by an observer at rest with respect to the scalar fluid
on the $xz$-plane,
where $z$ is the rotation axis and $x$ is an axis of the equatorial plane,
for the rotating radially excited solution (one node, $m=1$, $\kappa=0.05$)
at the turning point.
The white curve represents the location of the node.
}
\label{rprptm}
\end{figure}

The turning point solution, 
whose density and pressures are presented in Fig.~\ref{rprptm}, 
is the most compact boson star of the one node set of solutions. 
The energy density and radial pressure are highly concentrated 
in the inner shell, while the tangential pressure increases 
much more rapidly from the origin, 
assuming modestly negatives values 
outside the central parts of the shells.

\begin{figure}[h!]
\centering
\includegraphics[scale=0.58]{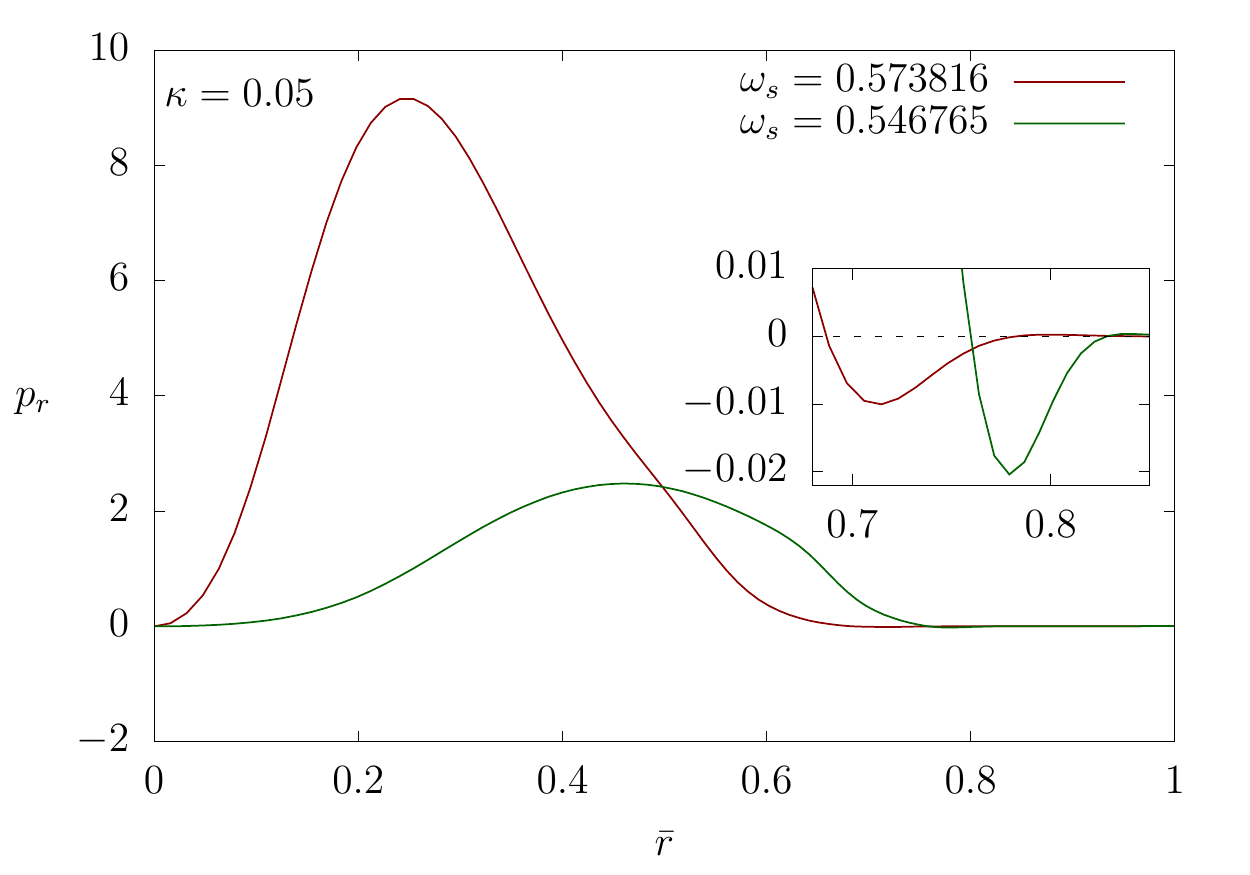}
\caption{
Radial pressure $p_r$
as measured by an observer at rest with respect to the scalar fluid
in the equatorial plane for the 
rotating radially excited solution at the turning point
and a solution on the upper branch close to it
(one node, $m=1$, $\kappa=0.05$).
The inset highlights that the radial pressure in the equatorial plane
can become slightly negative in the outer region of the outer shell.
}
\label{prme}
\end{figure}

The equatorial slice for the radial pressure $p_\bot$
is shown in more detail in Fig.~\ref{prme} 
for the solution at the turning point 
and a solution close to it located on the upper branch. 
The figure illustrates that,
when both rotation and radial excitation are present, 
some equilibrium configurations exhibit (albeit very sightly) 
a negative radial pressure in the equatorial plane,
an effect not seen in the nodeless rotating case. 

\subsection{Stability}
\label{sub_stab}

We here do not attempt to make a full stability analysis 
for these sets of solutions. 
Instead we will analyze the mass $M$
for a given particle number $Q$ together with the binding energy $E_b$
of the boson stars,
and determine which branch of solutions 
should more likely be the most stable one,
based on energetic grounds.
We thus follow similar lines of analysis as performed previously
for the spherically symmetric and rotating cases 
\cite{LEE1992251,PhysRevD.43.3895,PhysRevD.72.064002,PhysRevD.77.064025,PhysRevD.85.024045}.

Let us start with briefly recalling the stability of the fundamental rotating
boson stars in this model \cite{PhysRevD.85.024045}.
Here arguments from catastrophy theory have indicated, that there should be
two stable regions in the full set of nodeless rotating boson stars.
When comparing the set of rotating radially excited boson stars
with the fundamental one, one realizes, that for every allowed
particle number $Q$ of the radially excited solutions,
there is a fundamental solution with a lower mass.
This suggests that even the most stable radially excited solutions
can tunnel to a lower state, and thus will not be absolutely stable.
Let us therefore now focus on the relative stability of
the radially excited solutions,
looking for their most stable subset.

\begin{figure}[h!]
\begin{subfigure}{0.45\textwidth}
\includegraphics[scale=0.6]{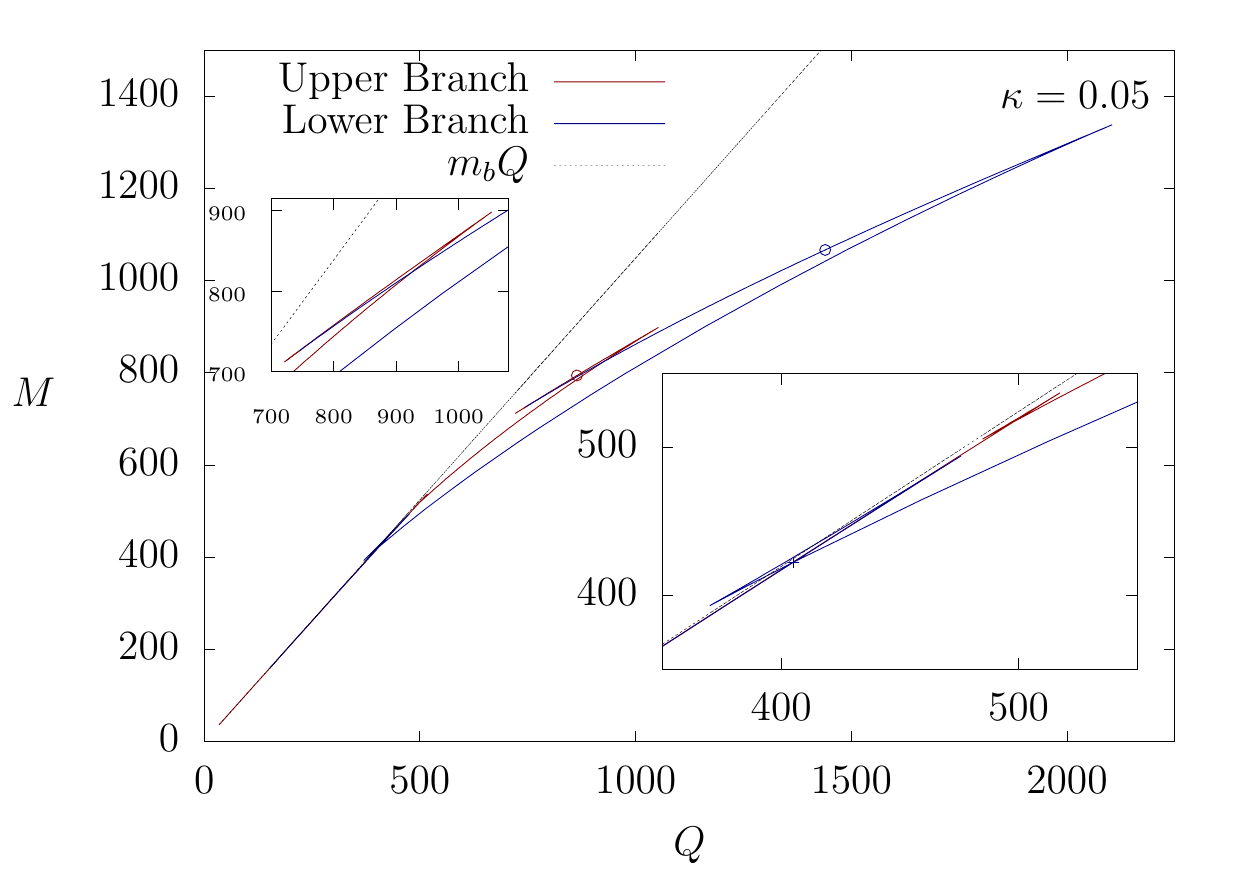}
\caption{}
\label{qm05}
\end{subfigure}
\begin{subfigure}{0.45\textwidth}
\includegraphics[scale=0.6]{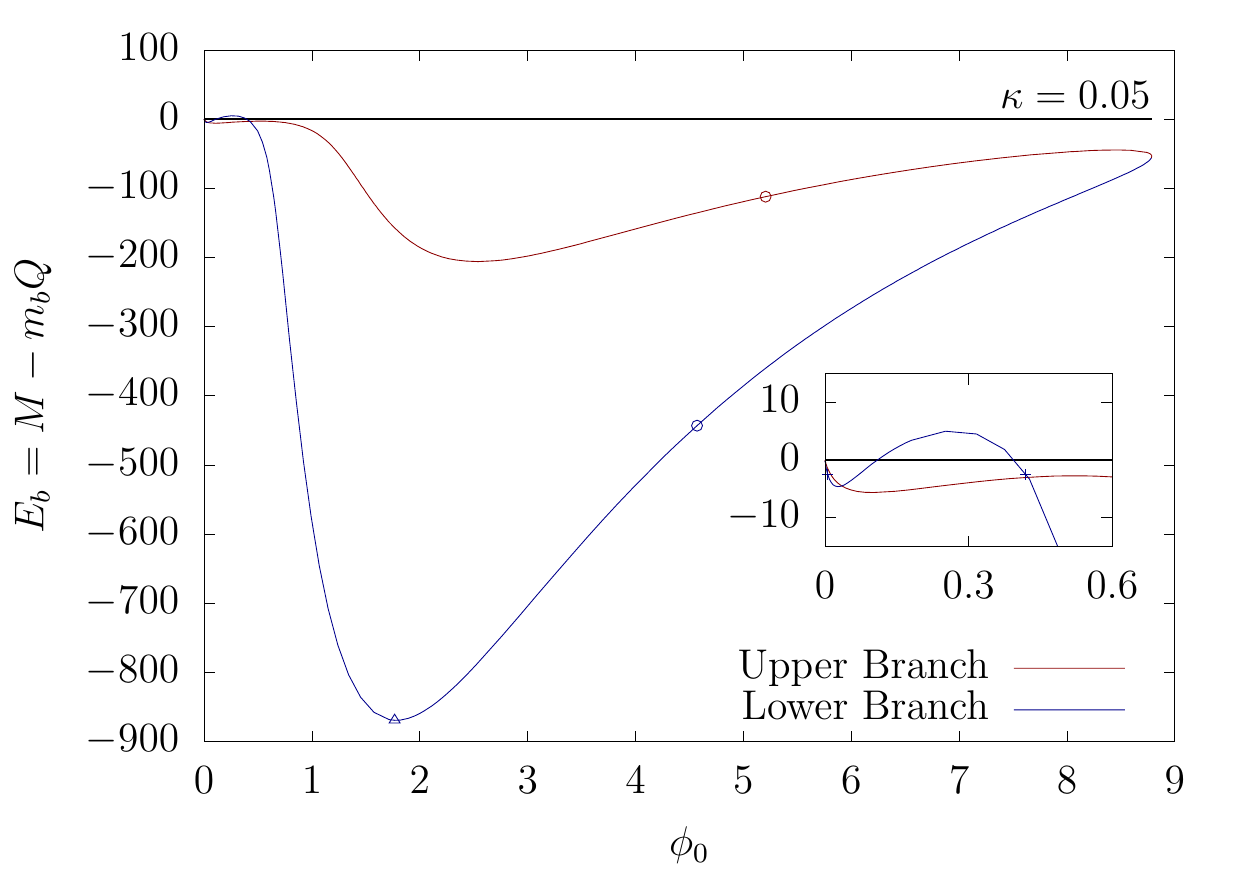}
\caption{}
\label{be05}
\end{subfigure}
\caption{
\emph{Left:} The mass $M$ of rotating radially excited boson stars ($m=1$)
with one node versus the particle number $Q$ ($\kappa=0.05$).
The circles indicate the onset and termination of the
presence of ergoregions.
The mass of $Q$ free particles, $m_bQ$ is shown for comparison.
\emph{Right:} The binding energy $E_b$ versus the value $\phi_0$, 
characterizing the boson field at the origin,
for the same set of solutions.
The triangle indicates the maximum mass solution,
the crosses indicate an $M$-$Q$ intersection point.}
\label{qmbe}
\end{figure}

Fig.~\ref{qm05} exhibits the mass $M$ versus the particle number $Q$ 
for the set of rotating radially excited boson stars
(with one node) for the coupling constant $\kappa=0.05$.
For better identification
the lower and upper branch are indicated by distinct colours.
Also shown is the straight line $M_{\rm free}=m_bQ$, indicating the mass of
$Q$ free bosons, and thus representing a boundary line for stability,
since solutions with a higher mass for a given particle number $Q$
could decay into $Q$ free particles.

In Fig.~\ref{qm05}
the upper and lower branch themselves contain a number of
subbranches, starting and ending at cusps, where
the mass and the particle number assume local or global maxima
and minima.
From previous mode analyses of spherically symmetric boson stars
it is known, that at these cusps the stability of the solutions
changes in the sense, that they acquire or lose an unstable mode.
Moreover, for a given particle number, the most stable
solution should correspond to the one with the lowest mass,
while higher mass solutions should be able to tunnel to the lowest
mass solution of a given particle number.

Inspection of the data and the figure then suggests that
the subset of solutions should be most stable,
which starts on the lower branch at the maximal value of 
$\omega_s$ from the vacuum, and reaches the cusp
at the global maximum of the mass and the particle number.
However, a small subset of this subset of solutions in the vicinity of
$Q\approx 500$ forms a swallow tail,
as seen in the lower right inset of the figure.
Thus in that region for a given value of the particle number
there are three (except at the two cusps, where there are two) 
solutions with different masses. 

Since the subset of solutions of the lower branch intersects itself
in the $M$-$Q$ diagram, when forming the swallow tail,
as marked by the cross in the lower right inset,
the subset of most stable solutions 
would simply jump at the intersection point 
from the lowest mass branch to the left of the intersection point
to the lowest mass branch to the right of the intersection point.
All other solution of the swallow tail will have
higher masses for a given particle number,
and should therefore be more unstable.
Note that some of the unstable solutions
possess even masses above the mass of the 
corresponding number of free particles.

According to these considerations, 
which can be made more stringent by applying
arguments from catastrophe theory
\cite{PhysRevD.43.3895,PhysRevD.85.024045},
all the remaining solutions on the lower branch 
beyond the maximal mass and all solutions
on the upper branch should be more unstable 
than this most stable subset.

However, there is another form of instability 
endangering the stability of regular rotating compact objects.
This is the ergoregion instability, discussed first for
boson stars in \cite{PhysRevD.77.124044}.
We indicate the onset and the termination
of the ergoregion for this set of boson stars in the figure
by the circles.
Clearly, here the ergoregion instability does not arise for the
most stable subset of solutions, but it
arises only for solutions, where stability has been lost
already.

Fig.~\ref{be05} shows the binding energy of this set of solutions
along the two branches defined by the value $\phi_0$,
characterizing the boson field at the origin. 
Here we note, that except for the small region of the swallow
tail around $Q\approx 500$, which is enlarged in the inset
and corresponds to small values of $\phi_0$,
the solutions on the lower branch are always more strongly bound
than the solutions on the upper branch.

This further confirms the above assessment, that the most
stable solutions reside on two parts of the lower branch.
At the global maximal value of the mass 
(indicated by the triangle in the figure),
which corresponds to the global maximal value of the particle number,
an unstable mode is expected to arise.
While the maximum mass configuration is most strongly bound,
it should also contain a zero mode, which should
turn into an unstable mode for configurations beyond the maximum mass
configuration.

Again, we emphasize, that there is the transition
in the swallow tail region around $Q\approx 500$,
where part of the lower branch would be more unstable.
We indicate by crosses in the inset of the figure the points, 
where this part of the lower branch 
intersects itself in the $M$-$Q$ diagram.
As noted above, all the solutions between these points,
should be more unstable.
The most stable set of solutions would directly
jump from the first cross to the next cross,
since mass, particle number and binding energy 
are the same for both configurations.
Note, that the particle number of the solutions
on the lower branch and on the upper branch for a given
value of $\phi_0$ can differ widely.

\section{Conclusions}
\label{s4}

Rotating boson stars are interesting compact objects,
formed by a (self-interacting) complex scalar field
coupled to gravity.
Their angular momentum $J$ is quantized,
representing an integer multiple $m$ of their particle number $Q$.
Their harmonic time-dependence is governed by the
boson frequency $\omega_s$, which assumes values in a finite
interval, bounded by the boson mass.

Here we have constructed solutions of rotating boson stars 
with rotational quantum number $m=1$,
which are radially excited.
The radial excitation is realized by the presence of a nodal surface,
where the scalar field vanishes.
We have found that
the presence of this radial excitation has profound consequences
for the set of boson star solutions.
They do not form a spiral, when their mass or particle number
are considered versus the frequency, as the set of fundamental
rotating boson star solutions does.
Instead they form a closed loop, starting from 
the maximal boson frequency at the vacuum solution,
reaching a minimal boson frequency and ending again
at the maximal boson frequency at the vacuum solution.

By considering the expansion of the scalar field at the origin,
one finds the expansion parameter $\phi_0$, 
which can be used to characterize the solutions.
In terms of $\phi_0$ one finds two branches of solutions,
extending between zero and a maximal value of $\phi_0$.
For each value of $\phi_0$ between the two limits
there are precisely two boson star solutions.
Thus the solutions form two branches of a loop,
a lower branch and an upper branch.
We have therefore labeled the solutions by their
value of $\phi_0$ and their location on one of
the branches.

The nodal surface of the rotating radially excited boson stars
divides the scalar fluid into two shells,
an inner shell and an outer shell.
The inner shell has the same qualitative properties in both branches
and is always centered on the equatorial plane.
In contrast, the outer shell's center
can be located at a wide range of values of the polar angle.
On the upper branch the outer shell
is closer to the equatorial plane,
while on the lower one it is nearer to the rotational axis.
These radially excited boson stars can thus possess
very different distributions of the scalar field.

Boson stars are anisotropic compact objects,
possessing different radial and tangential pressures.
Here we have also analyzed the energy density and the pressures
as seen by a comoving observer
for these rotating radially excited boson stars.
They all reflect the shell structure seen already for the
scalar field itself.
Interestingly, the most compact boson star is located
at the turning point of the loop,
i.e., at the maximal value of $\phi_0$. 

The scalar field of the boson stars can be decomposed
into a set of spherical harmonics with appropriate
radial functions.
Since the rotating boson stars studied here all possess 
a rotational quantum number $m=1$ and even parity,
the expansion is restricted accordingly.
The few lowest odd $\ell$ terms always suffice to obtain an 
excellent approximation of the solutions.
The decomposition of the most compact boson star
at the turning point of the loop
requires only two terms.
While these rotating boson stars always require
a superposition of modes,
one obtains (at least) two distinct superpositions
for each value of the boson frequency inbetween
the minimal and maximal values of the frequency.

We have also discussed the stability of the rotating boson stars,
following the reasoning of previous work.
This has led us to the conclusion that
for a given particle number and thus also
angular momentum, the solution with the lowest mass
and thus the most stable solution, will always
belong to the set of fundamental solutions,
i.e., be a nodeless solution.
Among the solutions with a radial excitation,
the most stable solutions reside on the lower branch
of solutions below the solution with the maximum mass.

\section{Acknowledgments}

We would like to acknowledge support by the DFG Research Training Group 1620
{\sl Models of Gravity} as well as by FP7, Marie Curie Actions, People,
International Research Staff Exchange Scheme (IRSES-606096).
BK gratefully acknowledges support 
from Fundamental Research in Natural Sciences
by the Ministry of Education and Science of Kazakhstan.

\appendix

\section{Field Equations}
\label{fea}

The Einstein field equations are solved as a boundary value problem. 
For simplicity we diagonalize them,
so that we work with five coupled elliptic PDEs written in the canonical form
\begin{align}
\label{efe1}
r^2\partial^2_rf+\partial^2_\theta f &= -\frac{1}{2}\frac{1}{fl}\Biggl[4\kappa r^2fl^2gU(\phi)-8\kappa m^2l^2g\omega^2\phi^2-16\kappa m\omega_srl^2g\omega\phi^2-8\kappa\omega_s^2r^2l^2g\phi^2-2\sin^2\theta l^2\omega^2 \nonumber \\
&+ 4rfl\partial_rf+2\cot\theta fl\partial_\theta f+f\partial_\theta f\partial_\theta l-2l\left(r^2\left(\partial_rf\right)^2+\left(\partial_\theta f\right)^2\right)+r^2f\partial_rf\partial_rl+4r\sin^2\theta l^2\omega\partial_r\omega \nonumber \\
&-2\sin^2\theta l^2\left(r^2\left(\partial_r\omega\right)^2+\left(\partial_\theta\omega\right)^2\right)\Biggr].
\end{align}

\begin{align}
\label{efe2}
r^2\partial^2_rl+\partial^2_\theta l &= -\frac{1}{2}\frac{1}{fl}\Biggl[8\kappa r^2l^3gU(\phi)+8\kappa\csc^2\theta m^2fl^2g\phi^2-8\kappa\frac{l^3g}{f}\phi^2\left(m^2\omega^2+2rm\omega_s\omega+r^2\omega_s^2\right)+6rfl\partial_rl \nonumber \\
&+4\cot^2\theta fl\partial_\theta l-f\left(r^2\left(\partial_rl\right)^2+\left(\partial_\theta l\right)^2\right)\Biggr].
\end{align}

\begin{align}
\label{efe3}
r^2\partial^2_rg+\partial^2_\theta g &= -\frac{1}{2}\frac{1}{fl}\Biggl[8\kappa m^2\csc^2\theta flg^2\phi^2-2\sin^2\theta\frac{l^2g}{f}\omega^2+8\kappa flg\left(\partial_\theta\phi\right)^2+4rfl\partial_rg-2\cot\theta fl\partial_\theta g \nonumber \\
&-2\frac{fl}{g}\left(r^2\left(\partial_r g\right)^2+\left(\partial_\theta g\right)^2\right)+f\left(r^2\partial_rl\partial_rg-\partial_\theta l\partial_\theta g\right)+2\frac{lg}{f}\left(\partial_\theta f\right)^2-2\cot\theta fg\partial_\theta l-3\frac{fg}{l}\left(\partial_\theta l\right)^2 \nonumber \\
&+2fg\partial_\theta^2l+4r\sin^2\theta\frac{l^2g}{f}\omega\partial_r\omega-2\sin^2\theta\frac{l^2g}{f}\left(r^2\left(\partial_r\omega\right)^2+2\left(\partial_\theta\omega\right)^2\right)\Biggr].
\end{align}

\begin{align}
\label{efe4}
r^2\partial^2_r\omega+\partial^2_\theta\omega &= -\frac{1}{2}\frac{1}{fl}\Biggl[8\kappa\csc^2\theta m^2flg\omega\phi^2-8\kappa r\csc^2\theta m\omega_sflg\phi^2-4fl\omega+4rfl\partial_r\omega+6\cot\theta fl\partial_\theta\omega \nonumber \\
&-4l\left(r^2\partial_rf\partial_r\omega+\partial_\theta f\partial_\theta\omega\right)+3f\left(r^2\partial_rl\partial_r\omega+\partial_\theta l\partial_\theta\omega\right)+4rl\omega\partial_rf-3rf\omega\partial_rl\Biggr].
\end{align}

\begin{align}
\label{eomphi}
r^2\partial^2_r\phi+\partial^2_\theta\phi &= -\frac{1}{2}\frac{1}{fl}\Biggl[-r^2l^2g\frac{\partial U\left(\phi\right)}{\partial\phi}-2\csc^2\theta m^2flg\phi+2\frac{l^2g}{f}\left(\omega m+r\omega_s\right)^2\phi+4rfl\partial_r\phi+2\cot\theta fl\partial_\theta\phi \nonumber \\
&+f\left(r^2\partial_rl\partial_r\phi+\partial_\theta l\partial_\theta\phi\right)\Biggr].
\end{align}

\section{Energy Density and Vierbein Components}

The energy density as measured by a particle comoving with the scalar fluid, 
given by $-T^0_0$ in eq.~(\ref{tab}) reads
\begin{align}
\rho=\frac{\phi^2\omega_s^2}{f}+\frac{f}{glr^2}\left[r^2\left(\partial_r\phi\right)^2+\left(\partial_\theta\phi\right)^2\right]+\frac{2\phi^2\omega_sm\omega}{fr}+\frac{\phi^2m^2}{lr^2\sin^2\theta}\left(\frac{l}{f}\omega^2\sin^2\theta-f\right)+U.
\end{align}

The eigenvector associated with the energy density 
is simply the four velocity of the fluid, $u^\mu=u^t(1,0,0,\Omega)$, 
such that $\Omega=u^\varphi/u^t=d\varphi/dt$ is its angular velocity. 
In extended form it reads
\begin{equation}
 \hat{e}^\mu_0=\left( \begin{array}{c}
\displaystyle
\frac{\sqrt{l}\sin\theta(m\omega+\omega_sr)}{\sqrt{-f(-lm^2\omega^2\sin^2\theta-2m\omega_sl\omega r\sin^2\theta-\omega_s^2lr^2\sin^2\theta+m^2f^2)}}\\[5pt]
0\\[5pt]
0\\[5pt]
\displaystyle
\frac{ml\omega^2\sin^2\theta+\omega_sl\omega r\sin^2\theta-mf^2}{\sqrt{-fl(-lm^2\omega^2\sin^2\theta-2m\omega_sl\omega r\sin^2\theta-\omega_s^2lr^2\sin^2\theta+m^2f^2)}r\sin\theta}
\end{array} \right). \refstepcounter{equation}\tag{\theequation}
\end{equation}

The eigenvector associated with the radial pressure reads
\[
 \hat{e}^\mu_1=\left( \begin{array}{c}
0\\[5pt]
\displaystyle
\frac{\sqrt{flg\left[r^2\left(\partial_r\phi\right)^2+\left(\partial_\theta\phi\right)^2\right]}r\partial_r\phi}{lg\left[r^2\left(\partial_r\phi\right)^2+\left(\partial_\theta\phi\right)^2\right]}\\[20pt]
\displaystyle
\frac{\sqrt{flg\left[r^2\left(\partial_r\phi\right)^2+\left(\partial_\theta\phi\right)^2\right]}\partial_\theta\phi}{lgr\left[r^2\left(\partial_r\phi\right)^2+\left(\partial_\theta\phi\right)^2\right]}\\[5pt]
0
\end{array} \right). \refstepcounter{equation}\tag{\theequation}
\]

The two eigenvectors associated with the tangential pressure are
given by
\[
 \hat{e}^\mu_2=\left( \begin{array}{c}
0\\[5pt]
\displaystyle
-\frac{\sqrt{flg\left[r^2\left(\partial_r\phi\right)^2+\left(\partial_\theta\phi\right)^2\right]}\partial_\theta\phi}{lg\left[r^2\left(\partial_r\phi\right)^2+\left(\partial_\theta\phi\right)^2\right]}\\[20pt]
\displaystyle
\frac{\sqrt{flg\left[r^2\left(\partial_r\phi\right)^2+\left(\partial_\theta\phi\right)^2\right]}\partial_r\phi}{lg\left[r^2\left(\partial_r\phi\right)^2+\left(\partial_\theta\phi\right)^2\right]}\\[5pt]
0
\end{array} \right), \refstepcounter{equation}\tag{\theequation}
\]

\[
 \hat{e}^\mu_3=\left( \begin{array}{c}
\displaystyle
\frac{-m\sqrt{f}}{\sqrt{(l\omega^2\sin^2\theta-f^2)m^2+2lm\omega\omega_sr\sin^2\theta+l\omega_s^2r^2\sin^2\theta}}\\[5pt]
0\\[5pt]
0\\[5pt]
\displaystyle
\frac{\omega_s\sqrt{f}}{\sqrt{(l\omega^2\sin^2\theta-f^2)m^2+2lm\omega\omega_sr\sin^2\theta+l\omega_s^2r^2\sin^2\theta}}\\
\end{array} \right). \refstepcounter{equation}\tag{\theequation}
\]

\bibliographystyle{ieeetr}
\bibliography{biblio}
\end{document}